\documentclass[12pt,english]{iopart}
\usepackage[T1]{fontenc}
\usepackage[latin2]{inputenc}
\usepackage{graphicx}

\providecommand{\tabularnewline}{\\}

\usepackage{iopams}
\usepackage{setstack}



\usepackage{babel}

\usepackage{babel}

\begin{document}

\title[Nonlinear]{Role of nonlinear vector meson interactions in hyperon stars.}

\author{ Ilona Bednarek, Ryszard Manka}

\address{Department of Astrophysics and Cosmology, Institute of Physics,\\
 University of Silesia, Uniwersytecka 4, PL-40-007 Katowice, Poland}

\maketitle

\begin{abstract}
The extended nonlinear model has been applied to construct neutron
star matter equation of state. In the case of neutron star matter
with non-zero strangeness the extension of the vector meson sector
by the inclusion of nonlinear mixed terms results in the
stiffening of the equation of state and accordingly in the higher
value of the maximum neutron star mass.
\end{abstract}

\section{Introduction}

The analysis of the role of strangeness in nuclear structure in
the aspect of multi-strange system is of great importance for both
nuclear physics and for astrophysics. In the latter case
understanding the properties of hyperon star is still a relevant
item. At the core of a neutron star the matter density ranges from
a few times the density of normal nuclear matter to about an order
of a magnitude higher,  at such densities exotic forms of matter
such as hyperons are expected to emerge. The appearance of these
additional degrees of freedom and their impact on a neutron star
structure have been the subject of extensive studies \cite{glen},
\cite{bema},
\cite{weber}, \cite{gal}.\\
The description of a neutron star interior is modelled on the
basis of the equation of state (EOS) of a dense nuclear system in
a neutron rich environment. In general the description of nuclear
matter is based on different models which can be grouped into
phenomenological and microscopic. Additionally each one of them
can be either relativistic or non-relativistic.   In a microscopic
approach the construction of the realistic model of
nucleon-nucleon (NN) interaction can be inspired by  the meson
exchange theory of nuclear forces. The parameters within the model
have to be adjusted to reproduce the experimental data for the
deuteron properties and NN scattering phase shifts
\cite{1995:Pethic}. The solution of the nuclear many-body problem
performed with the use of variational calculations for realistic
NN interactions (for example for the Argonne v14 or Urbana v14
potentials) saturate at the density $\sim 2 \times \rho_0$, where
$\rho_0$ denotes the saturation density  \cite{1981:lagaris,
1988:wiringa}. In order to obtain the correct description of
nuclear matter properties, namely the saturation density, binding
energy and compression modulus at the empirical values, a
phenomenological three-nucleon interaction has to be introduced.
Two-body forces, together with implemented three-body forces, help
providing the correct saturation point of symmetric nuclear matter
\cite{2002:Zuo}. The nuclear matter EOS calculated with the use of
the Brueckner--Hartree--Fock \cite{1991:bombaci}, \cite{1992:song}
approximation with the employed realistic two-nucleon interactions
(the Bonn and Paris potentials) also does not correctly reproduce
nuclear matter properties. Thus there are attempts to consider the
nuclear interaction problem in a relativistic formalism. The
relativistic version of the Brueckner--Hartree--Fock approximation
-- the Dirac--Brueckner--Hartree--Fock approach \cite{1992:Li} is
also based on realistic NN interactions. The nuclear EOS obtained
from the DBHF approach using the Bonn A potential  is soft at
moderate densities but become stiffer at higher densities
\cite{05:Dalen, 07:Klahn}. At densities up to 2-3 times nuclear
saturation density it is in agreement with constraints  from
heavy-ion collisions based on collective flow\cite{02:Danielewicz,
04:Stoicea} and kaon production \cite{06:Fuchs}.\\ The
relativistic approach to nuclear matter at the hadronic energy
scale developed by Walecka is very successful in describing a
variety of the ground-state properties of finite nuclei at, or
near the valley of stability and in predicting properties of
exotic nuclei with large neutron to proton excess. The standard
Walecka model \cite{walecka} comprises nucleons interacting
through the exchange of simulating medium range attraction
$\sigma$ mesons and $\omega$ mesons responsible for the short
range repulsion. Although this model properly described the
saturation point and the data for finite nuclei it has been
insufficient to properly describe the compression modulus of
symmetric nuclear matter at saturation density $K_{v}$ and the
proper density dependence in vector self-energy. The reproduced
value of the incompressibility coefficient obtained in the
original Walecka model gave large value of the order of $\sim 500$
MeV compared to the experimental results \cite{77:Youngblood}. The
nonlinear self-interactions of the scalar field (the cubic and
quartic terms) have been added in order to get the acceptable
value of the compression modulus $K_{v}$ \cite{bog77, bodmer}. Its
estimation made on the basis of recent experimental data point to
the range $230 \pm 10$ MeV \cite{04:Colo,
07:Garg, 07:Piekarewicz, 08:Chen}. \\
 The quartic vector self-interaction softens the high density component
of the EOS. This nonlinear term of the $\omega$ vector meson has
been used by Sugahara and Toki to construct TM1 and TM2 models
\cite{94:Sugahara}. However, models which satisfactorily reproduce
saturation properties of symmetric nuclear matter lead to
considerable differences in the case when density and asymmetry
dependence is included \cite{02:Furnstahl}. Isospin dependence of
the strong interactions between nucleons influences both physical
properties of nuclei and properties of infinite nuclear matter
\cite{04:Prakash}. The latter case includes mainly the description
of nuclear matter in high energy heavy-ion collisions and the
properties of neutron star matter. Thus, the proper model of
actual neutron star matter requires taking into consideration the
effect of neutron-proton asymmetry. This in turn leads to the
inclusion of the isovector meson $\rho$. The standard version of
the $\rho$ meson field introduction is of a minimal type without
any nonlinearities. This case has been further enlarged by the
nonlinear mixed isoscalar-isovector couplings which modify the
density dependence of the $\rho$ mean field and the energy
symmetry. Such an extension of the neutron star model has been
inspired by the paper \cite{hp} in which the authors show the
existence of a relationship between the neutron-rich skin of a
heavy nucleus and the properties
of a neutron star crust.\\
 The FSUGold model \cite{05:Todd} which lead to considerably softer
EOS. The meson sector of this model besides the linear terms of
the scalar and vector fields includes nonlinear isoscalar meson
self-interactions which soften the EOS of symmetric nuclear
matter, and the mixed isoscalar-isovector coupling which alters
the density dependence of the symmetry energy. Adding the mixed
isoscalar-isovector meson interaction term the FSUGold model
achieved acceptable results not only of the compression modulus
for symmetric nuclear matter ($K_{v}=230$ MeV) but the value of
the neutron skin in ${}^{208}$Pb of $R_{n}-R_{p}=0.21$ fm. The
main astrophysical prediction of this model is connected with
the value of the maximal neutron star mass which equals $M_{max}=1.72M_{\odot}$.\\
 They are the nonlinear vector self-interactions that are discussed
in this paper and the hadronic $SU(3)$ model which naturally
includes nonlinear scalar and vector interaction terms is likely
to be useful in the construction of models with nonzero
strangeness and with more accurate description of asymmetric
strangeness-rich neutron star matter \cite{sua}. The model
considered has been extended to include a broad spectrum of
nonlinear mixed vector meson couplings which stems from the very
special form of the vector meson sector. The choice of these
particular vector meson mixed interactions has been motivated by
the chiral SU(3) model. The main effect of such an extension of
the theory becomes evident when studying properties of neutron
star matter, especially the form of the EOS. The equations of
state for neutron star matter with hyperons considered in this
paper have shown considerable stiffening for higher densities.
Having obtained the equations of states the analysis of the
maximum achievable neutron star mass for given class of models can
be performed. Observational results limit the value of a neutron
star mass and thereby put constraints on the EOS of high density
nuclear matter. Recent observations of compact objects point to
the existence of a high maximum neutron star mass
\cite{08:Sagert}.  The most spectacular result obtained with the
Arecibo radio telescope for the neutron star-white dwarf binary
system predicted the largest neutron star mass ever reported
$M(PSRJ0751+1807) = 2.1\pm 0.2 M_{\odot} (1\sigma)$
\cite{05:Nice}. However, this result has been corrected by Nice
\cite{07:Klahn, 07:Nice}. The improved value of the orbital decay
and the detection of the Shapiro delay lead to the new value $1.26
M_{\odot}$ but with the estimated errors $1.12-1.30
M_{\odot}(1\sigma)$ and $0.98-1.53 M_{\odot}(2\sigma)$. Thus, this
case can not be used as a constraint on the EOS. But there are
another observations which indicate for high maximum mass of a
neutron star. These are among others the value of the radius ($R
> 12$ km)of the  isolated neutron star RX J1856.5-3754 \cite{04:Trumper} or  neutron
star mass in the low mass X-ray binary  (LMXB) 4U 1636-536
estimated at the value of $2.0\pm 0.1 M_{\odot}$ \cite{05:Barret}.
Another example of the LMXB is the neutron star in EXO 0748-676
which has been constrained by the detection of gravitational
redshift of certain absorption lines. This  combined with other
observational data lead to individually estimated mass and radius
of the star at the value of $M \geq 2.10 \pm 0.28 M_{\odot}$ and
$R \geq 13.8 \pm 1.8$ km  \cite{02:Cottam, 2006:Ozel}.

\section{Constituents of the model}

All calculations in this model have been done within the
theoretical framework of quantum hadrodynamics (QHD) and the
starting point is the nonlinear Walecka model which successfully
describes the properties of nuclear matter and finite nuclei
\cite{walecka}, \cite{walecka2}. Owing to substantial asymmetry of
neutron star matter models which include additional isospin
carrying terms can be used for its remarkably complete
description. Such models with Lagrangian functions supplemented by
isospin dependent nonlinear, mixed vector meson couplings have
been introduced and analyzed in papers by Piekarewicz et al.
\cite{bema}, \cite{hp}, \cite{2001:Piekarewicz}. In the model
considered there are couplings which relate the $\omega$ and
$\rho$ vector mesons with the $\phi$ meson and thus link the
asymmetry of the system with the strangeness content. The
inclusion of a broad spectrum of mixed nonlinear vector meson
couplings provides the possibility of modifying the high density
component of the EOS. However, the presence of additional terms in
the Lagrangian function requires the adjustment of new coupling
constants. This has been done by fitting
to the properties of nuclear matter.\\
 The presented model contains baryons and mesons as basic degrees
of freedom. The baryon-meson interactions and meson-meson interactions
are constructed on the basis of chiral SU(3) model \cite{sua}.\\

\subsection{Spin-0 meson fields}

Additional hadronic states are produced in neutron star interiors
in high density regime when the Fermi energy of nucleons exceeds the
hyperon masses. The hadronic SU(3) \cite{sua} model which includes
nonlinear scalar and vector interaction terms offers the possibility
of constructing a strangeness rich neutron star model and of providing
its detail description. The relevant degrees of freedom are hadrons
- composite fields constructed from quarks. The problem considered
is inspired by the chiral SU(3) model \cite{sua} in the nonlinear
realization which influences transformational properties of quarks.
The structure of hadrons is given in terms of their constituent quark
fields $q$ which can be split into left $q_{L}$ and right-handed
$q_{R}$ parts ($q=q_{L}+q_{R}$). They transform under $SU(3)_{L}\times SU(3)_{R}$
as \begin{equation}
q_{L}\rightarrow q_{L}^{'}=Lq_{L},\qquad q_{R}\rightarrow q_{R}^{'}=Rq_{R}.\end{equation}
 For the quarks of the nonlinear representation $\widetilde{q}$ the
following relation can be written \begin{equation}
q_{L}=u\widetilde{q}_{L}\qquad q_{R}=u^{\dag}\widetilde{q}_{R}\end{equation}
 where $\widetilde{q}_{L}$ and $\widetilde{q}_{R}(x)$ are left-
and right-handed components of the quark field $\widetilde{q}$ and
$u\equiv u(\pi(x))$ is given by \begin{equation}
u(\pi(x))=\sum_{a=1}^{8}\exp{\left(\frac{i}{2\sigma_{0}}\pi^{a}(x)\lambda_{a}\right)}\end{equation}
 with pseudoscalar mesons $\pi_{a}$ considered as parameters of the
symmetry transformation. They are identify with the octet of physical
pseudoscalar fields \cite{sua}.\\
 In general the meson content of the model consists of spin-0 and
spin-1 mesons. Introducing the matrix field $\Phi$ enables the collective
representation of the spin zero fields which in the matrix notation
can be written as $\Phi=\Sigma+i\Pi$. The $\Sigma$ and $\Pi$ mesons
can be transformed into nonlinearly transforming filds $X$ and $Y$
\begin{equation}
\Phi=\Sigma+i\Pi=u(X+iY)u,\end{equation}
 where $X$ is associated with the scalar nonet and $Y$ with the
pseudoscalar singlet which has to be added separately. The pseudoscalar
mesons appear as the parameters of the symmetry transformation. The
meson multiplets can be expanded in the basis of Gell-Mann matrices
thus $\Phi=\frac{1}{\sqrt{2}}T_{a}\phi_{a}$ where $T_{a}=\lambda_{a}$
are generators of $U(3)$ and $\lambda_{a}$ (a=1...8) are the Gell-Mann
matrices. It is convenient to introduce as a ninth matrix $\lambda_{0}=\sqrt{2/3}\, I$
($I$ is a unit matrix). The set $\{i\lambda_{a}\}_{a=0,\ldots8}$
constitutes a basis of the Lie algebra $u(3)$ of $U(3)$. \\
 The scalar fields in the basis of $U(3)$ generators are not mass
eigenstates. If the ideal mixing between the octet and singlet states
has been assumed then the octet state $\sigma_{8}$ and the singlet
state $\sigma_{0}$ are related to the ideal mixing states $\bar{\sigma}$
and $\bar{\zeta}$ of a scalar nonet through the orthogonal transformation
\begin{equation}
\left(\begin{array}{c}
\bar{\sigma}\\
\bar{\zeta}\end{array}\right)=\left(\begin{array}{cc}
cos\,\theta & sin\,\theta\\
-sin\,\theta & cos\,\theta\end{array}\right)\left(\begin{array}{c}
\sigma_{8}\\
\sigma_{0}\end{array}\right).\label{eq:scalar}\end{equation}
 Presenting the scalar multiplet as matrix the following form can
be obtained \begin{equation}
X=\left(\begin{array}{ccc}
\frac{(a_{0}^{0}+\bar{\sigma})}{\sqrt{2}} & a_{0}^{+} & \kappa^{+}\\
a_{0}^{-} & \frac{(-a_{0}^{0}+\bar{\sigma})}{\sqrt{2}} & \kappa^{0}\\
\kappa^{-} & \bar{\kappa^{0}} & \bar{\zeta}\end{array}\right).\end{equation}
 In the process of spontaneous chiral symmetry breaking $X$ acquires
the vacuum expectation value (VEV) $\equiv<X>$. As only components
proportional to $\lambda_{0}$ and the hypercharge $Y\sim\lambda_{8}$
are nonvanishing, $<X>$ takes the form $diag\{<\bar{\sigma}>,<\bar{\sigma}>,<\bar{\zeta}>\}$
and the following relations hold $f_{\pi}=\sqrt{<\bar{\sigma}>}$
and $f_{K}=(<\bar{\sigma}>+<\bar{\zeta}>)/\sqrt{2}$ where $f_{\pi}$
and $f_{K}$ are the pseudoscalar decay constants.\\
 Making references to the Walecka model the following transformation
should be done \begin{eqnarray}
\bar{\sigma} & = & \sigma+<\bar{\sigma}>\\
\bar{\zeta} & = & \sigma^{\ast}+<\bar{\zeta}>\nonumber \end{eqnarray}
 where $\sigma$ and $\sigma^{\ast}$ denotes fields in the Walecka
model.\\
 In this paper the mean field approach serves as a method for solving
the many body problem. In this approximation meson fields are separated
into classical mean field values and quantum fluctuations, which are
not included in the ground state. Thus, for the ground state of homogeneous
infinite nuclear matter quantum fields operators are replaced by their
classical expectation values $s_{0}$ and $s_{0}^{\ast}$ \begin{eqnarray}
\sigma & = & \tilde{\sigma}+s_{0}\\
\sigma^{\ast} & = & \tilde{\sigma}^{\ast}+s_{0}^{\ast}.\nonumber \end{eqnarray}

\subsection{Spin-1 meson fields}

The spin-1 mesons are given by two octets of vector and axial vector
fields. These fields similarly as in the case of spin zero mesons
also can be written in a compact form \begin{equation}
l_{\mu}(r_{\mu})=\frac{1}{2}(V_{\mu}\pm A_{\mu})=\frac{1}{2\sqrt{2}}\sum_{a=0}^{a=8}(v_{\mu}^{a}\pm a_{\mu}^{a})\lambda^{a},\end{equation}
 where $l_{\mu}$ and $r_{\mu}$ corresponds to left and right-handed
gauge fields, respectively and $V_{\mu}$ and $A_{\mu}$ denote the
vector and axial vector nonets respectively.\\
 The physical isoscalar $\omega$ and $\phi$ meson fields stem
from the pure singlet $v_{\mu}^{0}$ and octet $v_{\mu}^{8}$ isoscalar
states and can be obtained assuming ideal mixing, namely that $\phi_{\mu}$
is a pure $\bar{s}s$ state. This come down to the following relations
\begin{eqnarray}
\phi_{\mu} & = & \frac{1}{\sqrt{3}}(\sqrt{2}v_{\mu}^{0}+v_{\mu}^{8})\\
\omega_{\mu} & = & \frac{1}{\sqrt{3}}(v_{\mu}^{0}-\sqrt{2}v_{\mu}^{8}).\nonumber \end{eqnarray}
 Accordingly the vector meson multiplet can be expressed explicitly
in terms of physical fields as \begin{equation}
V_{\mu}=\left(\begin{array}{ccc}
\frac{(\omega_{\mu}+\rho_{\mu}^{0})}{\sqrt{2}} & \rho_{\mu}^{+} & K^{\ast+}\\
\rho_{\mu}^{-} & \frac{(\omega_{\mu}+\rho_{\mu}^{0})}{\sqrt{2}} & K^{\ast0}\\
K^{\ast-} & \overline{K}^{\ast0} & \phi_{\mu}\end{array}\right).\label{VM}\end{equation}

\subsection{Baryon fields}

Nonlinearly transforming baryon fields can be written as \begin{equation}
B_{L}=u^{\dag}\Psi_{L}u\qquad B_{R}=u\Psi_{R}u^{\dag}\end{equation}
 where $\Psi_{L}$ and $\Psi_{R}$ are the left and right-handed parts
of the baryon field in the linear representation. Baryon fields that
enter the model are grouped into $3\times3$ traceless matrix $\mathcal{B}$
\begin{equation}
\mathcal{B}=\left(\begin{array}{ccc}
\frac{1}{\sqrt{6}}\Lambda+\frac{1}{\sqrt{2}}\Sigma^{0} & \Sigma^{+} & p\\
\Sigma^{-} & \frac{1}{\sqrt{6}}\Lambda-\frac{1}{\sqrt{2}}\Sigma^{0} & n\\
\Xi^{-} & \Xi^{0} & -\frac{2}{\sqrt{6}}\Lambda\end{array}\right).\label{baryon}\end{equation}

\section{The model}

The dynamics of the system has been described in terms of the Lagrangian
function which in its most general form can be given as a sum of two
basic parts representing meson and baryon sectors which are directly
related by the term defining the baryon-meson interaction \begin{equation}
\mathcal{L}=\mathcal{L}_{kin}+\mathcal{L}_{M}+\mathcal{L}_{B}+\mathcal{L}_{BM}.\end{equation}
 Additionally the kinetic term $\mathcal{L}_{kin}$ for both baryon
and meson fields has been included.

\subsection{The meson sector}

The chiral SU(3) theory provides the basis for calculations made in
this paper, however, the relativistic mean field approach to the description
of the static, uniform nuclear matter leads to useful generalization
about the model considered. In general, the meson sector includes
contributions from spin zero and spin one mesons but in the mean field
approximation vacuum expectation value of the pseudoscalar and axial
meson fields vanishes and these fields do not enter to the Lagrangian
function. In the result meson field Lagrangian $\mathcal{L}_{M}$
embodies only the parts for scalar and vector mesons \begin{equation}
\mathcal{L}_{M}=\mathcal{L}_{S}+\mathcal{L}_{V}.\end{equation}
 The scalar part of the Lagrangian function $\mathcal{L}_{M}$ includes
the potential terms \begin{equation}
\mathcal{L}_{S}=-k_{1}(I_{2})^{2}+k_{2}I_{4}+2k_{3}I_{3},\end{equation}
 which are given in the form of chiral invariants defined as \begin{equation}
I_{1}=Tr(X),\,\, I_{2}=Tr(X)^{2},\,\, I_{3}=detX.\end{equation}
 The vector meson Lagrangian function presented in this model is the
sum of a mass term and vector meson self-interaction terms up to fourth
order in the fields \begin{equation}
\mathcal{L}_{V}=\frac{1}{2}m_{v}^{2}Tr(V_{\mu}V^{\mu})+\mathcal{L}_{VV}.\label{lagV}\end{equation}
 The second part of (\ref{lagV}) can be written in the form of invariants
\begin{equation}
\mathcal{L}_{VV}=\frac{1}{4}c\,(Tr(V^{\mu}V_{\mu}))^{2}+\frac{1}{2}d\, Tr[(V^{\mu}V_{\mu})^{2}]+\frac{1}{16}f[Tr(V^{\mu})]^{4},\label{lavec}\end{equation}
 where $V_{\mu}$ is the vector meson multiplet. In order to split
the mass degeneracy for the meson nonet the following chiral invariant
has to be added \begin{equation}
\mathcal{L}_{MV}=\frac{1}{4}\mu Tr[V_{\mu\nu}V^{\mu\nu}X^{2}].\end{equation}
 This together with the kinetic energy term, which will be introduced
in section 3.3, leads to the following terms for different vector
mesons \begin{eqnarray}
 &  & -\frac{1}{4}[1-\mu\frac{\bar{\sigma}^{2}}{2}](V_{\rho}^{\mu\nu})^{2}-\frac{1}{4}[1-\frac{1}{2}\mu(\frac{\bar{\sigma}^{2}}{2}+k^{2})](V_{K^{*}}^{\mu\nu})^{2}\\
 &  & -\frac{1}{4}[1-\mu\frac{\bar{\sigma}^{2}}{2}](V_{\omega}^{\mu\nu})^{2}-\frac{1}{4}[1-\mu k^{2}](V_{\phi}^{\mu\nu}).\nonumber \end{eqnarray}
 As the coefficients are not equal unity the vector meson fields have
to be renormalized by the factor $Z_{\omega}^{-1}(\bar{\sigma})=1-\mu\bar{\sigma}^{2}/2$.
The mass terms of the vector mesons differ from the mean mass $m_{V}$
by the renormalization factor and the following result can be obtained
\begin{equation}
m_{\omega}^{2}=m_{\rho}^{2}=Z_{\omega}(\sigma_{0})m_{V}^{2};\quad m_{K^{\ast}}^{2}=Z_{K^{\ast}}(\zeta_{0})m_{V}^{2};\quad m_{\phi}^{2}=Z_{\phi}(\zeta_{0})m_{V}^{2}\end{equation}
 where the constants $m_{V}$ $\mu$ and $k$ are fixed to give the
correct $\omega$ and $\phi$ masses and $\sigma_{0}=<\bar{\sigma}>|_{n_{B}=0}$
and $\zeta_{0}=<\bar{\zeta}>|_{n_{B}=0}$, $n_{B}$ denotes the baryon
number density.\\
 The baryonic part of the Lagrangian can be written as \begin{equation}
\mathcal{L}_{B}=Tr(\bar{\mathcal{B}}i\gamma^{\mu}D_{\mu}\mathcal{B}),\end{equation}
 where $\mathcal{B}$ is a $3\times3$ traceless hermitian matrix
given by relation (\ref{baryon}) and $D_{\mu}$ denotes the covariant
derivative of $\mathcal{B}$ which is defined as \begin{equation}
D_{\mu}\mathcal{B}=\partial_{\mu}\mathcal{B}+i[\Gamma_{\mu},\mathcal{B}]\end{equation}
 with $\Gamma_{\mu}$ defined as \begin{equation}
\Gamma_{\mu}=-\frac{i}{2}[u^{\dag}(\partial_{\mu}+ig_{v}l_{\mu})u+u(\partial_{\mu}+ig_{v}r_{\mu})u^{\dag}].\label{gamma}\end{equation}

\subsection{Baryon-meson interaction}

Using the notation introduced by \cite{sua} the very general SU(3)
structure for the baryon-meson interaction can be written as \begin{eqnarray}
\mathcal{L}_{BM} & = & -\sqrt{2}g_{8}^{W}(\alpha_{W}[\bar{\mathcal{B}}\mathcal{O}\bar{B}W]_{F}+(1-\alpha_{W})[\bar{\mathcal{B}}\mathcal{O}\bar{\mathcal{B}}W]_{D})\\
 & - & g_{1}^{W}\frac{1}{\sqrt{3}}Tr(\bar{\mathcal{B}}\mathcal{O}\mathcal{B})TrW,\nonumber \end{eqnarray}
 where $W$ denotes general meson field and \begin{equation}
[\bar{\mathcal{B}}\mathcal{O}\mathcal{B}W]_{F}:=Tr(\bar{\mathcal{B}}\mathcal{O}W\mathcal{B}-\bar{\mathcal{B}}\mathcal{O}\mathcal{B}W)\end{equation}
 \begin{equation}
[\bar{\mathcal{B}}\mathcal{O}\mathcal{B}W]_{D}:=Tr(\bar{\mathcal{B}}\mathcal{O}W\mathcal{B}+\bar{\mathcal{B}}\mathcal{O}\mathcal{B}W)-\frac{2}{3}Tr(\bar{\mathcal{B}}\mathcal{O}\mathcal{B})TrW.\end{equation}
 Different forms of the presented above interaction terms result from
differences in the Lorentz structure.\\
 For the nonlinear realization of chiral symmetry the antisymmetric
(F-type) and symmetric (D-type) interaction terms of baryons not only
with spin-1 but with spin-0 mesons as well are allowed.\\
 In the case of baryon-scalar meson interaction $W=X,\mathcal{O}=\mathcal{I}$,
for vector meson $W=V_{\mu},\mathcal{O}=\gamma_{\mu}$, for axial
vector mesons $W=\mathcal{A_{\mu}},\mathcal{O}=\gamma_{\mu}\gamma_{5}$
and for pseudo-scalar mesons $W=u_{\mu},\mathcal{O}=\gamma_{\mu}\gamma_{5}$.\\
 Baryon-spin zero meson interaction is indispensable for the construction
of baryon mass terms. Masses of the whole baryon multiplet are generated
spontaneously by the vacuum expectation values (VEV) of the non-strange
and strange scalar fields. When the nucleon mass depends on the strange
condensate $<\bar{\zeta}>$ the parameters $g_{1}^{S},g_{8}^{S}$
and $\alpha_{S}$ enable baryon masses to be fitted to their experimental
values \begin{eqnarray}
m_{N}(\bar{\sigma},\bar{\zeta}) & = & m_{0}-\frac{1}{3}g_{8}^{S}(4\alpha_{S}-1)(\sqrt{2}\bar{\zeta}-\bar{\sigma})\\
m_{\Lambda}(\bar{\sigma},\bar{\zeta}) & = & m_{0}-\frac{2}{3}g_{8}^{S}(\alpha_{S}-1)(\sqrt{2}\bar{\zeta}-\bar{\sigma})\nonumber \\
m_{\Sigma}(\bar{\sigma},\bar{\zeta}) & = & m_{0}+\frac{2}{3}g_{8}^{S}(4\alpha_{S}-1)(\sqrt{2}\bar{\zeta}-\bar{\sigma})\nonumber \\
m_{\Xi}(\bar{\sigma},\bar{\zeta}) & = & m_{0}+\frac{1}{3}g_{8}^{S}(2\alpha_{S}+1)(\sqrt{2}\bar{\zeta}-\bar{\sigma}),\nonumber \end{eqnarray}
 where $m_{0}$ is determined by two meson field condensates and is
given by \begin{equation}
m_{0}=g_{1}^{S}\frac{\sqrt{2}<\bar{\sigma}>+<\bar{\zeta}>}{\sqrt{3}}.\end{equation}
 However, the assumption that $\alpha_{S}=1$ and $g_{1}^{S}=\sqrt{6}g_{8}^{S}$
leads to the model in which nucleon mass depends only on the nonstrange
condensate $<\bar{\sigma}>$. In this case the coupling constance
between the baryons and the two scalar condensates are related to
the additive quark model. Then nucleon mass depends only on the non-strange
condensate $<\bar{\sigma}>$, and only one coupling constant is needed
to reproduce the correct value of the nucleon mass. To obtain the
correct masses of the remaining baryons an explicit symmetry breaking
term has to be added.\\
 The explicit symmetry breaking term presented in the paper by
\cite{sua} has been used \begin{equation}
\mathcal{L}_{\Delta m}=-m_{1}Tr(\bar{\mathcal{B}}\mathcal{B}-\bar{\mathcal{B}}\mathcal{B}S)-m_{2}Tr(\bar{\mathcal{B}}S\mathcal{B})\end{equation}
 where $S_{b}^{a}=1/3(\sqrt{3}(\lambda_{8})_{a}^{b}-\delta_{b}^{a})$
and the parameters $g_{8}^{S},m_{1}$ and $m_{2}$ are used to determine
baryon masses: \begin{eqnarray}
m_{N} & = & -g_{N\sigma}<\bar{\sigma}>\\
m_{\Xi} & = & -\frac{1}{3}g_{N\sigma}<\bar{\sigma}>-\frac{2}{3g_{N\sigma}}\sqrt{2}\bar{<\zeta>}+m_{1}+m_{2}\nonumber \\
m_{\Lambda} & = & -\frac{2}{3}g_{N\sigma}<\bar{\sigma}>-\frac{1}{3g_{N\sigma}}\sqrt{2}\bar{<\zeta>}+\frac{m_{1}+2m_{2}}{3}\nonumber \\
m_{\Sigma} & = & -\frac{2}{3}g_{N\sigma}<\bar{\sigma}>-\frac{1}{3g_{N\sigma}}\sqrt{2}\bar{<\zeta>}+m_{1}.\nonumber \end{eqnarray}
 Considering the case when nucleon mass depends only on non-strange
condensate and once again making references to the Walecka model the
relation for the baryon masses can be expressed as \begin{equation}
m_{B}(\bar{\sigma},\bar{\zeta})=m_{B}(\sigma,\sigma^{\ast})=m_{B}-g_{B\sigma}\sigma-g_{B\sigma^{\ast}}\sigma^{\ast},\end{equation}
 where the terms $g_{B\sigma}\sigma$ and $g_{B\sigma^{\ast}}\sigma^{\ast}$
represent the modification of baryon masses due to the medium.\\
 The interaction of baryons with spin-1 mesons can be construct
analogously with the baryon spin-0 meson interaction. For the case
of pure F-type coupling ($\alpha_{V}=1$) the assumption $g_{1}^{V}=\sqrt{6}g_{8}^{V}$
(the strange vector field $\phi_{\mu}\sim\overline{s}\gamma_{\mu}s$
does not couple to nucleon) can be made. As it has been stated in
the mean field approach the VEV of axial mesons are zero thus taking
into account only vector mesons the Lagrangian describing baryon-vector
meson interaction is given by \begin{equation}
\mathcal{L}_{BV}=-\sqrt{2}g_{8}^{V}\left(Tr\bar{\mathcal{B}}\gamma^{\mu}[V_{\mu}^{8},\mathcal{B}]+Tr\bar{\mathcal{B}}\gamma^{\mu}\mathcal{B}\cdot TrV_{\mu}^{1}\right)\label{BV}\end{equation}
 After insertion of the matrix (\ref{VM}) to equation (\ref{BV})
and using all the facts concerning the construction of the baryon
mass the following form of the Lagrangian function can be written
\begin{equation}
\mathcal{L}_{MB}=Tr(\bar{\mathcal{B}}(i\gamma^{\mu}D_{\mu}-m_{B}(\bar{\sigma},\bar{\zeta}))\mathcal{B})\end{equation}
 where $m_{B}$ denotes the mass of the baryon octet in the chiral
limit and the covariant derivative of $\mathcal{B}$ is given by \begin{equation}
D_{\mu}\mathcal{B}=\partial_{\mu}\mathcal{B}+i[\Gamma_{\mu},\mathcal{B}],\end{equation}
 with the connection $\Gamma_{\mu}$.\\
 In the vector meson sector the couplings to the strange baryons
determined from the symmetry relations read \begin{eqnarray}
g_{N\omega} & = & (4\alpha_{V}-1)g_{8}^{V}\\
g_{\Lambda\omega} & = & \frac{2}{3}(5\alpha_{V}-2)g_{8}^{V}\qquad g_{\Lambda\phi}=-\frac{\sqrt{2}}{3}(2\alpha_{V}+1)g_{8}^{V}\nonumber \\
g_{\Sigma\omega} & = & 2\alpha_{V}g_{8}^{V}\qquad g_{\Sigma\phi}=-2\sqrt{2}(2\alpha_{V}-1)g_{8}^{V}\nonumber \\
g_{\Xi\omega} & = & (2\alpha_{V}-1)g_{8}^{V}\qquad g_{\Xi\phi}=-2\sqrt{2}\alpha_{V}g_{8}^{V}\nonumber \end{eqnarray}
 Taking $\alpha_{V}=1$ the coupling constants related to the additive
quark model can be obtained \begin{equation}
g_{\Lambda\omega}=g_{\Sigma\omega}=g_{\Xi\omega}=\frac{2}{3}g_{N\omega}=2g_{8}^{V}\qquad g_{\lambda\phi}=g_{\Sigma\phi}=\frac{g_{\Xi\phi}}{2}=\frac{\sqrt{2}}{3}g_{N\omega}.\label{veccoup}\end{equation}
 Similar symmetry relations can be obtained for the coupling constants
in the scalar sector. These however are not used in the presented
model. The couplings of baryons with the scalar mesons are determined
from the experimentally estimated value of the $\Lambda$ central
potential.

\subsection{Kinetic terms}

The explicit form of the baryon and meson kinetic terms are given
by \begin{equation}
\mathcal{L}_{kin}=iTr(\bar{\mathcal{B}}\gamma_{\mu}D^{\mu}\mathcal{B})+\frac{1}{2}Tr(D_{\mu}XD^{\mu}X)-\frac{1}{4}Tr(V_{\mu\nu}V^{\mu\nu})\end{equation}
 where $D_{\mu}$ denotes the covariant derivative \begin{equation}
D_{\mu}=\partial_{\mu}+i[\Gamma_{\mu},]\label{covder}\end{equation}
 and $\Gamma_{\mu}$ is given by relation (\ref{gamma}). The chirally
invariant kinetic term for spin-1 mesons, in the case of vector
mesons, is $V_{\mu\nu}=D_{\mu}V_{\nu}-D_{\nu}V_{\mu}$. The
covariant derivative $D_{\mu}$ is given by relation
(\ref{covder}).

\section{The effective model}

Vector mesons and thereby vector densities are the decisive
factors that contribute to the EOS of dense matter in neutron star
interiors. Thus, attention is focused on the construction of a
model which includes wide spectrum of nonlinear couplings between
vector meson fields. This allows one to perform a systematic
analysis
of their influence on the high density EOS.\\
 As has been stated in previous section theoretical description
of strange hadronic matter requires the extension of the nonlinear
Walecka model by the inclusion of baryons of the lowest SU(3) flavor
octet. In order to describe the strongly attractive $\Lambda\Lambda$
interaction two additional meson fields, the scalar meson $f_{0}(975)$
denoted as $\sigma^{\ast}$ and the vector meson $\phi(1020)$ have
been introduced \cite{Schaffner}.\\
 Thus, in the scalar meson sector besides non-strange $\sigma$
meson the hidden-strange $\sigma^{\ast}$ scalar meson is included
whereas, in the case of vector mesons $\omega$, $\rho$ and $\phi$
mesons are comprised.\\
 Summing-up the Lagrangian function for the system consists of
a baryonic part which includes the full octet of baryons together
with terms describing interaction of baryons with scalar and vector
mesons and a mesonic part. The mesonic part contains also additional
interactions between mesons which mathematically express themselves
as supplementary, nonlinear terms in the Lagrangian function. Considering
individual constituents of the model which have been described in
the previous paragraph the very general form of Lagrangian function
can be written \begin{equation}
{\mathcal{L}_{BM}}=\sum_{B}\bar{\psi}_{B}i\gamma^{\mu}D_{\mu}\psi_{B}-\sum_{B}m_{B}(\sigma,\sigma^{\ast})\overline{\psi}_{B}\psi_{B}+\mathcal{L}_{M},\end{equation}
 where baryon fields $\Psi_{B}^{T}=(\psi_{N},\psi_{\Lambda},\psi_{\Sigma},\psi_{\Xi})$
are composed of the following isomultiplets \cite{glen}: 
\[
\begin{array}{cc}
{\displaystyle \Psi_{N}={\psi_{p} \choose \psi_{n}},} & {\displaystyle \Psi_{\Lambda}=\psi_{\Lambda},}\\
\\{\displaystyle \Psi_{\Sigma}=\left(\begin{array}{c}
\psi_{\Sigma^{+}}\\
\psi_{\Sigma^{0}}\\
\psi_{\Sigma^{-}}\end{array}\right),} & {\displaystyle \Psi_{\Xi}={\psi_{\Xi^{0}} \choose \psi_{\Xi^{-}}},}\end{array}\]
 $D_{\mu}$ is the covariant derivative of baryons which in terms
of $\omega_{\mu},\rho_{\mu}^{a}$ and $\phi_{\mu}$ fields is given
by \begin{equation}
D_{\mu}=\partial_{\mu}+ig_{\omega B}\omega_{\mu}+ig_{\phi B}\phi_{\mu}+ig_{\rho B}I_{3B}\tau^{a}\rho_{\mu}^{a}.\label{devcovs}\end{equation}
 The meson part of the Lagrangian function \begin{eqnarray}
\mathcal{L}_{M}=\frac{1}{2}\partial_{\mu}\sigma\partial^{\mu}\sigma+\frac{1}{2}\partial_{\mu}\sigma^{\ast}\partial^{\mu}\sigma^{\ast}-U_{eff}(\sigma,\sigma^{\ast},\omega_{\mu},\rho_{\mu}^{a},\phi_{\mu})\nonumber \\
-\frac{1}{4}\Omega_{\mu\nu}\Omega^{\mu\nu}-\frac{1}{4}R_{\mu\nu}^{a}R^{a\mu\nu}-\frac{1}{4}\Phi_{\mu\nu}\Phi^{\mu\nu}\end{eqnarray}
 includes the field tensors $\Omega_{\mu\nu},\Phi_{\mu\nu}$ and $R_{\mu\nu}^{a}$
defined as \begin{equation}
\Omega_{\mu\nu}=\partial_{\mu}\omega_{\nu}-\partial_{\nu}\omega_{\mu}\hspace{0.5cm}\Phi_{\mu\nu}=\partial_{\mu}\phi_{\nu}-\partial_{\nu}\phi_{\mu}\label{eq:fieldwf}\end{equation}
 \begin{equation}
R_{\mu\nu}^{a}=\partial_{\mu}\rho_{\nu}^{a}-\partial_{\nu}\rho_{\mu}^{a}.\label{eq:fieldr}\end{equation}
 All meson interaction terms are collected in the potential function
$U_{eff}(\sigma,\sigma^{\ast},\omega_{\mu},\rho_{\mu}^{a},\phi_{\mu})$
which can be written as a sum of linear and nonlinear parts \begin{equation}
\mathcal{U}_{eff}(\sigma,\sigma^{\ast},\omega_{\mu},\rho_{\mu}^{a},\phi_{\mu})=\mathcal{U}_{lin}(\sigma,\sigma^{\ast},\omega_{\mu},\rho_{\mu}^{a},\phi_{\mu})+\mathcal{U}_{nl}(\sigma,\sigma^{\ast},\omega_{\mu},\rho_{\mu}^{a},\phi_{\mu}).\label{poteff}\end{equation}
 The linear scalar and vector meson part of the potential takes the
form \begin{eqnarray}
\mathcal{U}_{lin}(\sigma,\sigma^{\ast},\omega_{\mu},\rho_{\mu}^{a},\phi_{\mu}) & = & \frac{1}{2}m_{\sigma}^{2}\sigma^{2}+\frac{1}{2}m_{\sigma^{\ast}}^{2}\sigma^{\ast2}-\frac{1}{2}m_{\omega}^{2}(\omega_{\mu}\omega^{\mu})\\
 & - & \frac{1}{2}m_{\rho}^{2}(\rho_{\mu}^{a}\rho^{\mu a})-\frac{1}{2}m_{\phi}^{2}(\phi_{\mu}\phi^{\mu}),\nonumber \end{eqnarray}
 whereas its nonlinear part is given by \begin{eqnarray}
\mathcal{U}_{nl}(\sigma,\sigma^{\ast},\omega_{\mu},\rho_{\mu}^{a},\phi_{\mu})=\frac{1}{3}g_{3}\sigma^{2}+\frac{1}{4}g_{4}\sigma^{4}+\frac{1}{4}(c+2d+\frac{1}{4}f)(\phi_{\mu}\phi^{\mu})^{2}+\\
+\frac{1}{4}(c+d)(\rho_{\mu}^{a}\rho^{\mu a})^{2}+\frac{1}{2}c(\rho_{\mu}^{a}\rho^{\mu a})(\phi_{\nu}\phi^{\nu})+\frac{1}{2\sqrt{2}}f(\omega_{\mu}\phi^{\mu})(\phi_{\nu}\phi^{\nu})+\nonumber \\
+\frac{1}{2}(c(\rho_{\mu}^{a}\rho^{\mu a})+3d(\rho_{\mu}^{a}\rho^{\mu a})+c(\phi_{\mu}\phi^{\mu})+\frac{3}{2}f(\phi_{\mu}\phi^{\mu})(\omega_{\nu}\omega^{\nu}))+\nonumber \\
+\frac{1}{\sqrt{2}}f(\phi_{\mu}\omega^{\mu})(\omega_{\nu}\omega^{\nu})+\frac{1}{4}(c+d+f)(\omega_{\mu}\omega^{\mu})^{2}.\nonumber \end{eqnarray}
 The potential function $\mathcal{U}_{eff}(\sigma,\sigma^{\ast},\omega_{\mu},\rho_{\mu}^{a},\phi_{\mu})$
includes subsets of non-strange and strangeness rich matter and this
can be written with the use of the following notation \begin{equation}
\mathcal{U}_{eff}(\sigma,\sigma^{\ast},\omega_{\mu},\rho_{\mu}^{a},\phi_{\mu})=\mathcal{U}_{i,S=0}+\mathcal{U}_{i,S\neq0},\end{equation}
 where $i$ denotes linear and non-linear parts of the potential (\ref{poteff}).
If strangeness bearing components are not taking into account in the
description of matter and only nucleons are considered, the Lagrangian
density function constructed on the basis of the Walecka model can
be retained which in the case of symmetric matter comprises contributions
coming from nucleons, $\sigma$ and $\omega_{\mu}$ mesons. As neutron
star matter is highly asymmetric one the inclusion of isovector-vector
meson $\rho_{\mu}^{a}$ become indispensable. Finally making reference
to the extended Walecka model with nonlinear scalar and vector meson
self-interaction terms the Lagrangian function can be presented as
follows \begin{eqnarray}
\mathcal{L} & = & \bar{\psi}_{N}(i\gamma^{\mu}D_{\mu}-(m_{N}(\sigma)-g_{\sigma N}\sigma))\psi_{N}+\frac{1}{2}\partial_{\mu}\sigma\partial^{\mu}\sigma-\nonumber \\
 & - & \frac{1}{4}\Omega_{\mu\nu}\Omega^{\mu\nu}-\frac{1}{4}R_{\mu\nu}^{a}R^{a\mu\nu}+\frac{1}{2}m_{\omega}^{2}\omega_{\mu}\omega^{\mu}\nonumber \\
 & + & \frac{1}{2}m_{\rho}\rho_{\mu}^{a}\rho^{a\mu}-U_{eff,S=0}(\sigma,\omega_{\mu},\rho_{\mu}^{a})\end{eqnarray}
 where \[
\psi_{N}=\left(\begin{array}{c}
\psi_{p}\\
\psi_{n}\end{array}\right)\]
 is the nucleon field, $D_{\mu}$ denotes the covariant derivative
which now reduces to \begin{equation}
D_{\mu}=\partial_{\mu}+ig_{\omega N}\omega_{\mu}+ig_{\rho N}I_{3N}\tau^{a}\rho_{\mu}^{a}.\end{equation}
 The nucleon mass is denoted by $m_{N}(\sigma)$ $(N=n,p)$ whereas
$m_{i}$ $(i=\sigma,\omega,\rho)$ are masses assigned to the meson
fields, $R_{\mu\nu}^{a}$ and $\Omega_{\mu\nu}$ again denotes the
field tensors and are given by (\ref{eq:fieldwf}) and (\ref{eq:fieldr}).\\
 Collecting all the meson self-interactions and the mixed $\omega-\rho$
and $\sigma-\rho$ terms in the model considered a nonlinear part
of the potential $U_{eff,S=0}(\sigma,\omega_{\mu},\rho_{\mu}^{a})$
can be specified \begin{eqnarray}
U_{nl,S=0}(\sigma,\omega_{\mu},\rho_{\mu}^{a})=-\frac{\kappa}{3!}\sigma^{3}-\frac{\lambda}{4!}\sigma^{4}+c_{3}(\omega_{\mu}\omega^{\mu})^{2}\\
+\frac{\zeta}{4!}g_{\rho N}^{4}(\rho_{\mu}^{a}\rho^{\mu a})^{2}+\Lambda_{V}(g_{\omega N}g_{\rho N})^{2}(\omega_{\mu}\omega^{\mu})(\rho_{\mu}^{a}\rho^{\mu a})+\Lambda_{4}(g_{\sigma N}g_{\rho N})^{2}\sigma^{2}(\rho_{\mu}^{a}\rho^{\mu a}).\nonumber \end{eqnarray}
 More detailed analysis of the vector meson influence on the high
density EOS needs to consider different types of nonlinear
vector meson couplings.\\
 The very general form of the vector meson Lagrangian
(\ref{lavec})
 includes the potential which is defined by coupling constants that appeared
in the vector meson Lagrangian (\ref{lavec}). Since the coupling
constants of the mixed vector meson interactions are known very
poorly, if ever, there are still important uncertainties in the
analysis of their influence on the form of the EOS. In order to
study the importance of individual parts of the potential it is
necessary to find connections between the coupling constants $c$,
$d$ and $f$ and the couplings in the vector part of the nonlinear
potential $U_{nl}(\sigma,\omega_{\mu},\rho_{\mu}^{a})$.\\
 The results reveal the following relations \begin{equation}
c=\frac{1}{2}(3c_{3}-3c_{3}\beta-2g_{\rho}^{2}g_{\omega}^{2}\Lambda_{V})\label{eq:par1}\end{equation}
 \begin{equation}
d=\frac{1}{2}(-c_{3}+c_{3}\beta+2g_{\rho}^{2}g_{\omega}^{2}\Lambda_{V})\label{eq:par2}\end{equation}
 \begin{equation}
f=c_{3}\beta.\label{eq:par3}\end{equation}
 In the paper by Serot et. al \cite{serot} the acceptable values
of the parameters entering the very general nonlinear model have
been introduced. In accordance with the estimations made in the
paper \cite{serot} a new parameter $\beta$ which takes the value
in the range <0;1> can be introduced. This parameter makes
possible to carry out more systematic analysis and to include
wider class of models that have been constructed on the basis of
the invariants presented in (\ref{lavec}). The considered cases
lead to models and results which already have been reported.
Incorporating the relations (\ref{eq:par1}-\ref{eq:par3}) the
vector potential  can be constructed \begin{eqnarray}
\mathcal{U}_{V} & = & \frac{1}{4}c_{3}\left((\rho_{\mu}^{a}\rho^{\mu a})^{2}(1-\beta)+(\omega_{\mu}\omega^{\mu})^{2}\right)+\label{UV}\\
 & + & (\omega_{\mu}\omega^{\mu})\left(\frac{3}{4}c_{3}(\phi_{\mu}\phi^{\mu})+\frac{1}{2}(g_{\rho}g_{\omega})^{2}\Lambda_{V}(2(\rho_{\mu}^{a}\rho^{\mu a})-(\phi_{\mu}\phi^{\mu})\right)+\nonumber \\
 & + & (\rho_{\mu}^{a}\rho^{\mu a})\left(\frac{3}{4}c_{3}(\phi_{\mu}\phi^{\mu})-\frac{3}{4}\beta\phi_{\mu}\phi^{\mu}-\frac{1}{2}(g_{\rho}g_{\omega})^{2}\Lambda_{V}\phi_{\mu}\phi^{\mu}\right)+\nonumber \\
 & + & \frac{1}{8}(\phi_{\mu}\phi^{\mu})^{2}\left(c_{3}(1-\frac{1}{2}\beta)+2(g_{\omega}g_{\rho})^{2}\Lambda_{V}\right)\nonumber \\
 & + & \frac{3}{4}c_{3}(\rho_{\mu}^{a}\rho^{\mu a})(\phi_{\mu}\phi^{\mu})(1-\beta)+\frac{1}{2\sqrt{2}}\beta c_{3}(\phi_{\nu}\phi^{\nu})(\phi_{\mu}\omega^{\mu})+\nonumber \\
 & + & \frac{3}{4}c_{3}(\phi_{\mu}\phi^{\mu})(\omega_{\mu}\omega^{\mu})+\frac{1}{\sqrt{2}}\beta c_{3}(\phi_{\mu}\omega^{\mu})(\omega_{\nu}\omega^{\nu}).\nonumber \end{eqnarray}
 Detailed analysis of the presented above potential can be made considering
different cases.
\begin{description}
\item [{{1.}}] Nonstrange matter $(S=0)$. This case includes variety
of models and leads directly to the well-known results presented in
the literature.\\
 The potential (\ref{UV}) reduces to the most general form for
non-strange matter \begin{equation}
\mathcal{U}_{V}=\frac{1}{4}c_{3}(\rho_{\mu}^{a}\rho^{\mu a})^{2}\left(1-\beta\right)+\frac{1}{4}c_{3}(\omega_{\mu}\omega^{\mu})^{2}+(g_{\rho}g_{\omega})^{2}\Lambda_{V}(\rho_{\mu}^{a}\rho^{\mu a})(\omega_{\mu}\omega^{\mu}).\end{equation}
 The analysis starts with the case $\beta=0$ and $\Lambda_{V}=0$.
\begin{equation}
\mathcal{U}_{V}=\frac{1}{4}c_{3}(\rho_{\mu}^{a}\rho^{\mu a})^{2}+\frac{1}{4}c_{3}(\omega_{\mu}\omega^{\mu})^{2}\end{equation}
 includes contributions from both isoscalar and isovector components.
The last one is of special interest for the the asymmetric neutron
star matter.\\
 The case with $\Lambda_{V}\neq0$\\
 \begin{equation}
\mathcal{U}_{V}=\frac{1}{4}c_{3}(\rho_{\mu}^{a}\rho^{\mu a})^{2}+\frac{1}{4}c_{3}(\omega_{\mu}\omega^{\mu})^{2}+(g_{\rho}g_{\omega})^{2}\Lambda_{V}(\rho_{\mu}^{a}\rho^{\mu a})(\omega_{\nu}\omega^{\nu}).\end{equation}
 This form of the potential function incorporates additional vector
meson mixed term.\\
 The case with $\beta=1$ and $\Lambda_{V}\neq0$ \begin{equation}
\mathcal{U}_{V}=\frac{1}{4}c_{3}(\omega_{\mu}\omega^{\mu})^{2}+(g_{\rho}g_{\omega})^{2}\Lambda_{V}(\rho_{\mu}^{a}\rho^{\mu a})(\omega_{\nu}\omega^{\nu})\end{equation}
 This is the model studied by Piekarewicz et al. \cite{hp}, \cite{2001:Piekarewicz}\\
 The case with $\beta=1$ and $\Lambda_{V}=0$ \begin{equation}
\mathcal{U}_{V}=\frac{1}{4}c_{3}(\omega_{\mu}\omega^{\mu})^{2}\end{equation}
 The nonlinear Walecka model is retained in this situation.
\item [{{2.}}] Matter with nonzero strangeness $(S\neq0)$. \\
 The case with $\beta=0$ \begin{eqnarray}
\mathcal{U}_{V} & = & \frac{1}{4}c_{3}(\rho_{\mu}^{a}\rho^{\mu a})^{2}+\frac{3}{4}c_{3}(\rho_{\mu}^{a}\rho^{\mu a})(\phi_{\nu}\phi^{\nu})+\frac{1}{8}c_{3}(\phi_{\mu}\phi^{\nu})^{2}\\ \nonumber
 &  & +\frac{3}{4}c_{3}(\phi_{\mu}\phi^{\mu})(\omega_{\nu}\omega^{\nu})+\frac{1}{4}(g_{\rho}g_{\omega})^{2}\Lambda_{V}(\phi_{\mu}\phi^{\mu})^{2}\\ \nonumber
 &  & +\frac{1}{4}c_{3}(\omega_{\mu}\omega^{\mu})^{2}-\frac{1}{2}(g_{\rho}g_{\omega})^{2}\Lambda_{V}(\rho_{\mu}^{a}\rho^{\mu a})(\phi_{\nu}\phi^{\nu})\\  \nonumber
 &  & +(g_{\rho}g_{\omega})^{2}\Lambda_{V}(\rho_{\mu}^{a}\rho^{\mu a})(\omega_{\nu}\omega^{\nu})-\frac{1}{2}(g_{\rho}g_{\omega})^{2}\Lambda_{V}(\phi_{\mu}\phi^{\mu})(\omega_{\nu}\omega^{\nu})
 \end{eqnarray}
 The case with $\beta=0$ and $\Lambda_{V}=0$\begin{eqnarray}
\mathcal{U}_{V} & = & \frac{1}{4}c_{3}(\rho_{\mu}^{a}\rho^{\mu a})^{2}+\frac{3}{4}c_{3}(\rho_{\mu}^{a}\rho^{\mu a})(\phi_{\nu}\phi^{\nu})\nonumber \\
 & + & \frac{1}{8}c_{3}(\phi_{\mu}\varphi^{\mu})^{2}+\frac{3}{4}c_{3}(\phi_{\mu}\phi^{\mu})(\omega_{\nu}\omega^{\nu})\end{eqnarray}
 The case with $\beta=1$ and $\Lambda_{V}\neq0$ \begin{eqnarray}
\mathcal{U}_{V} & = & \frac{1}{16}c_{3}(\phi_{\mu}\phi^{\mu})^{2}++\frac{1}{2\sqrt{2}}c_{3}(\phi_{\mu}\phi^{\mu})(\phi_{\nu}\omega^{\nu})+\frac{3}{4}c_{3}(\phi_{\mu}\phi^{\mu})(\omega_{\nu}\omega^{\nu})\nonumber \\
 & + & \frac{1}{\sqrt{2}}c_{3}(\phi_{\mu}\omega^{\mu})(\omega_{\nu}\omega^{\nu})+\frac{1}{4}c_{3}(\omega_{\mu}\omega^{\mu})^{2}-\frac{1}{2}(g_{\rho}g_{\omega})^{2}\Lambda_{V}(\rho_{\mu}^{a}\rho^{\mu a})(\phi_{\nu}\phi^{\nu})\nonumber \\
 & + & \frac{1}{4}(g_{\rho}g_{\omega})^{2}\Lambda_{V}(\phi_{\mu}\phi^{\mu})^{2}\nonumber \\
 & + & (g_{\rho}g_{\omega})^{2}\Lambda_{V}\rho_{\mu}^{2}(\omega_{\mu}\omega^{\mu})-\frac{1}{2}(g_{\rho}g_{\omega})^{2}\Lambda_{V}(\phi_{\mu}\phi^{\mu})(\omega_{\nu}\omega^{\nu})\end{eqnarray}
 The case with $\beta=1$ and $\Lambda_{V}=0$ \begin{eqnarray}
\mathcal{U}_{V} & = & \frac{1}{16}c_{3}(\phi_{\mu}\phi^{\mu})^{2}+\frac{1}{2\sqrt{2}}c_{3}(\phi_{\mu}\phi^{\mu})(\phi^{\nu}\omega_{\nu})+\\
 & + & \frac{3}{4}c_{3}(\phi_{\mu}\phi^{\mu})(\omega_{\nu}\omega^{\nu})+\frac{1}{\sqrt{2}}c_{3}(\phi_{\mu}\omega^{\mu})(\omega_{\nu}\omega^{\nu})+\frac{1}{4}c_{3}(\omega_{\mu}\omega^{\mu})^{2}.\nonumber \end{eqnarray}

\end{description}
Summing up the vector meson potential in general can include a
wide variety of nonlinear terms. The presence of nonzero
strangeness led to very distinct division of the constructed
models. This has been done with the use of the parameter $\beta$.
For nucleonic matter ($S=0$) both cases $\beta = 0$ and $\beta =
1$ have been analyzed. The latter one is of special importance as
it enables the comparison of the results with those obtained by
Piekarewicz et al. \cite{2001:Piekarewicz}. However, in the case
of nonzero strangeness important reduction has been done. Models
constructed for the case $\beta=1$ lead to very soft equation of
state and consequently to very low value of the maximum neutron
star masses. This is not interested according to results of recent
observations which point towards larger masses. Such masses are
especially concerned about the stiffening of the EOS at
sufficiently high densities. This stiffening of the equation of
state has become a big issue particularly for strangeness rich
matter. Thus, further analysis of the influence of the nonlinear
vector meson interactions for strangeness rich matter will
concentrate on the case with $\beta=0$.

\subsection{Mean Field Approximation}

The system considered, has been assumed to be isotropic, infinite
matter in its ground state. To investigate properties of infinite
nuclear matter, the mean field approximation has been adopted. The
symmetries of infinite nuclear matter simplify the model to a great
extent. The translational and rotational invariance claimed that the
mean fields of all the vector fields vanish. Only the time-like components
of the neutral vector mesons have a non-vanishing expectation value.
Owing to parity conservation, the vacuum expectation value of pseudoscalar
fields vanish ($<\pi_{a}>=0$). Meson fields have been separated into
classical mean field values and quantum fluctuations, which are not
included in the ground state. Thus, for the ground state of homogeneous
infinite nuclear matter quantum fields operators are replaced by their
classical expectation values. Hence, baryons move independently in
the mean meson fields which generate themselves self-consistently
by baryons. The resulting field equations for the mean field approximation
have a reduced, simpler form \begin{equation}
m_{\sigma}^{2}s_{0}+g_{3}s_{0}^{2}+g_{4}s_{0}^{3}=\sum_{B}g_{\sigma B}m_{eff,B}^{2}S(m_{eff,B},k_{F,B}),\end{equation}
 \begin{equation}
m_{eff\omega}^{2}w_{0}=\sum_{B}g_{\omega B}n_{B},\end{equation}
 \begin{equation}
m_{eff\rho}^{2}r_{0}=\sum_{B}g_{\rho B}I_{3B}n_{B},\end{equation}
 \begin{equation}
m_{\sigma^{\ast}}^{2}s_{0}^{\ast}=\sum_{B}g_{\sigma^{\ast}B}m_{eff,B}^{2}S(m_{eff,B}),\end{equation}
 \begin{equation}
m_{eff\varphi}^{2}f_{0}=\sum_{B}g_{\varphi B}n_{B},\end{equation}
 where $s_{0},w_{0},r_{0},s_{0}^{\ast}$ and $f_{0}$ are the classical
mean field values of the meson fields and $m_{i,eff}$ are effective
masses assigned to $\omega$, $\rho$ and $\phi$ meson fields.
The effective masses are given by the relations \begin{equation}
m_{eff,\omega}^{2}=m_{\omega}^{2}+3c_{3}w_{0}^{2}+2\Lambda_{V}(g_{\omega}g_{\rho})^{2}r_{0}^{2}+2(\frac{3}{4}c_{3}-\frac{1}{2}\Lambda_{V}(g_{\omega}g_{\rho})^{2})f_{0}^{2}\end{equation}
 \begin{equation}
m_{eff\rho}^{2}=m_{\rho}^{2}+c_{3}r_{0}^{2}+2\Lambda_{V}(g_{\omega}g_{\rho})^{2}w_{0}^{2}+2(\frac{3}{4}c_{3}-\frac{1}{2}\Lambda_{V}(g_{\omega}g_{\rho})^{2})w_{0}f_{0}^{2}\end{equation}
 \begin{equation}
m_{eff\varphi}^{2}=m_{\varphi}^{2}+3c_{3}r_{0}^{2}+2\Lambda_{V}(g_{\omega}g_{\rho})^{2}w_{0}^{2}+4(\frac{3}{4}c_{3}-\frac{1}{2}\Lambda_{V}(g_{\omega}g_{\rho})^{2})w_{0}^{2}f_{0}\end{equation}
 The function $S(m_{eff,B},k_{F,B})$ is expressed with the use of
an integral \begin{equation}
S(m_{eff,B},k_{F,B})=\frac{2J_{B}+1}{2\pi^{2}}\int_{0}^{k_{FB}}\frac{m_{eff,B}}{\sqrt{k^{2}+m_{eff,B}}}k^{2}dk,\label{funkcjaS}\end{equation}
 where $J_{B}$ and $I_{3B}$ are the spin and isospin projection
of baryon B, $k_{F,B}$ is the Fermi momentum of species $B$, $n_{B}$
denotes the baryon number density. The presence of the $\sigma^{\ast}$
and $\phi$ meson fields provides new potential terms to the Dirac
equation which now takes the form \begin{equation}
(i\gamma^{\mu}\partial_{\mu}-m_{eff,B}-g_{\omega B}\gamma^{0}w_{0}-g_{\rho B}I_{3B}\gamma^{0}\tau^{3}r_{0}-g_{\varphi B}\gamma^{0}f_{0})\psi_{B}=0\end{equation}
 with $m_{eff,B}$ being the effective baryon mass generated by the
baryon and scalar fields interaction and defined as\\
 \begin{equation}
m_{eff,B}=m_{B}-(g_{\sigma B}s_{0}+g_{\sigma^{\ast}B}s_{0}^{\ast}).\label{masseff}\end{equation}
 In order to calculate the energy density and pressure of the nuclear
matter the energy-momentum tensor $T_{\mu\nu}$ which is given by
the relation \begin{equation}
T_{\mu\nu}\equiv\frac{\partial\mathcal{L}}{\partial(\partial_{}\mu\phi_{i})}\partial^{\nu}\phi_{i}-\eta_{\mu\nu}\mathcal{L}\label{energymomtensor}\end{equation}
 have to be used. In equation (\ref{energymomtensor}) $\phi_{i}$
denotes both boson and fermion fields.\\
 The energy density $\epsilon$ equals $<T_{00}>$ whereas the
pressure $P$ is related to the statistical average of the trace of
the spatial component $T_{ij}$ of the energy-momentum tensor.
Calculations done for the considered model lead to the following
explicit formula for the energy density and pressure:
\begin{eqnarray}
\epsilon=\frac{1}{2}m_{\omega}^{2}w_{0}^{2}+\frac{3}{4}c_{3}w_{0}^{4}+\frac{1}{2}m_{\rho}^{2}r_{0}^{2}
+\frac{1}{2}m_{\phi}^{2}f_{0}^{2}+\frac{1}{2}m_{\sigma^{\ast}}^{2}s_{0}^{\ast 2}+\epsilon_{B}+\\
\nonumber
+3\Lambda_{V}(g_{\rho}g_{\omega})^{2}w_{0}^{2}r_{0}^{2}+\frac{3}{4}r_{0}^{4}+
U(s_{0})+3\left(\frac{1}{8}c_{3}+\frac{1}{4}\Lambda_{V}(g_{\rho}g_{\omega})^{2}
\right)f_{0}^{4}+\\ \nonumber +3\left(
\frac{3}{4}c_{3}-\frac{1}{2}\Lambda_{V}(g_{\rho}g_{\omega})^{2}\right)f_{0}^{2}r_{0}^{2}
\label{energy1}\end{eqnarray}
 with $\epsilon_{B}$ given by\\
 \begin{equation}
\epsilon_{B}=\sum_{B}\frac{2}{\pi^{2}}\int_{0}^{k_{F,B}}k^{2}dk\sqrt{k^{2}+(m_{B}-g_{\sigma B}s_{0}-g_{\sigma^{\ast}}s_{0}^{\ast})^{2}},\end{equation}
 \begin{eqnarray}
P=\frac{1}{2}m_{\rho}r_{0}^{2}+\frac{1}{2}m_{\omega}w_{0}^{2}+\frac{1}{4}c_{3}(w_{0}^{4}+r_{0}^{4})+\frac{1}{2}m_{\phi}^{2}f_{0}^{2}-\frac{1}{2}m_{\sigma^{\ast}}^{2}s_{0}^{\ast 2}-\\
\nonumber -U(s_{0})
+\Lambda_{V}(g_{\rho}g_{\omega})^{2}w_{0}^{2}r_{0}^{2}
+\left(\frac{3}{4}c_{3}-\frac{1}{2}\Lambda_{V}(g_{\rho}g_{\omega})^{2}\right)f_{0}^{2}(w_{0}^{2}+r_{0}^{2})+\\
\nonumber
+\left(\frac{1}{8}c_{3}+\frac{1}{4}\Lambda_{V}(g_{\rho}g_{\omega})^{2}
\right)f_{0}^{4}+P_{B} \label{psressure1}\end{eqnarray}
 \begin{equation}
P_{B}=\sum_{B}\frac{1}{3\pi^{2}}\int_{0}^{k_{F,B}}\frac{k^{4}dk}{\sqrt{(k^{2}+m_{B}-g_{\sigma B}s_{0}-g_{\sigma^{\ast}}s_{0}^{\ast})^{2}}}.\end{equation}
 The obtained form of the EOS
determines the physical state and composition of matter at high
densities. In order to construct the neutron star model through
the entire density span it is necessary to add the EOS,
characteristic for the inner and outer core, relevant to lower
densities. Thus, a more complete and more realistic description of
a neutron star requires taking into consideration not only the
interior region of a neutron star, but also its remaining layers.
In these calculations the composite EOS has been constructed by
joining together the EOS of the neutron rich matter core region
and neutron star crust. The inner crust is a region which spans
from the neutron drip point to the inner boundary separating the
solid crust from the homogeneous core \cite{09:JunXu} \cite{Jun
Xu}. Since the density drops steeply near the surface of a neutron
star, these layers do not contribute significantly to the total
mass of a neutron star. The inner neutron rich region up to
density $\rho\sim10^{13}\, g\, cm^{-3}$ influences decisively the
neutron star structure and evolution.

\section{The equilibrium conditions and composition of stellar matter.}

The ground state of a neutron star is thought to be the question of
equilibrium dependence on the baryon and electric-charge conservation.
Neutrons are the principal components of a neutron star when the density
of matter is comparable to the nuclear density. For higher densities
it is the equilibrium of the process \begin{equation}
p+e^{-}\leftrightarrow n+\nu_{e}\end{equation}
 which establishes the relation between chemical potentials \begin{equation}
\mu_{p}+\mu_{e}=\mu_{n}+\mu_{\nu_{e}}.\end{equation}
 Thus, realistic neutron star models describe electrically neutral
high density matter being in $\beta$ equilibrium. The latter condition
implies the presence of leptons. It is expressed by adding the Lagrangian
of free leptons \begin{equation}
\mathcal{L}_{L}=\sum_{f=e,\mu}\overline{\psi}_{f}(i\gamma^{\mu}\partial_{\mu}-m_{f})\psi_{f}.\end{equation}
 Neutrinos are neglected here since they leak out from a neutron star,
whose energy diminishes at the same time. After electron chemical
potential $\mu_{e}$ has reached the value equal to the muon mass,
muons start to appear. Equilibrium with respect to the reaction \begin{equation}
e^{-}\leftrightarrow\mu^{-}+\nu_{e}+\bar{\nu}_{\mu}\end{equation}
 is assured when $\mu_{\mu}=\mu_{e}$ (setting $\mu_{\nu_{e}}=\mu_{\bar{\nu}_{\mu}}=0$).
The appearance of muons reduces number of electrons and also affects
the amount of the protons. \\
 Additional hadronic states are produced in neutron star interiors
at sufficiently high densities when hyperon in-medium energy equals
their chemical potential. The onset of hyperon formation depends on
the attractive hyperon-nucleon interaction. The higher the density
the more various hadronic species are expected to populate. They can
be formed both in leptonic and baryonic processes. In the latter the
relevant strong interaction processes that establish the hadron population
in neutron star matter e.g.: \begin{equation}
\Lambda+\Lambda\leftrightarrow\Xi+N\hspace{8mm}(Q=25MeV)\end{equation}
 or \begin{equation}
\Sigma+N\rightarrow\Lambda+N\hspace{8mm}(Q=80MeV)\end{equation}
 are Pauli blocked. Taking into account the energy released in these
reactions (denoted as $Q$)  it is likely that $\Sigma$
hyperons do not appear in neutron star matter. The chemical potentials
of neutron star components are related in such a way that the chemical
equilibrium in stellar matter can be achieved. The requirement of
charge neutrality and equilibrium under the week processes in the
instance of strangeness rich matter \begin{equation}
B_{1}\rightarrow B_{2}+f+\bar{\nu}_{f}\hspace{0.5cm}B_{2}+f\rightarrow B_{1}+\nu_{f}\end{equation}
 leads to the following relations: \begin{eqnarray}
\sum_{i}\left(n_{B_{i}^{+}}+n_{f^{+}}\right)=\sum_{i}\left(n_{B_{i}^{-}}+n_{f^{-}}\right)\\
\mu_{i}=b_{i}\mu_{n}+q_{i}\mu_{f}\nonumber \end{eqnarray}
 where $b_{i}$ is the baryon number of particle $i$, $q_{i}$ is
its charge, $f$ stands for leptons $f=e,\mu$, $B_{i}$ denotes baryons
and $\mu_{\nu_{f}}=0$. The conditions mentioned above result in the
relations between chemical potentials and constrain the species fractions
in the stellar interior when taking into consideration the baryon
octet and leptons included in this model: \begin{eqnarray}
\mu_{p}=\mu_{\Sigma^{+}}=\mu_{n}-\mu_{e}\hspace{10mm}\mu_{\Lambda}=\mu_{\Sigma^{0}}=\mu_{\Xi^{0}}=\mu_{n}\\
\mu_{\Sigma^{-}}=\mu_{\Xi^{-}}=\mu_{n}+\mu_{e}\hspace{10mm}\mu_{\mu}=\mu_{e}.\nonumber \end{eqnarray}

\section{Parameters}

Nuclear matter can be defined as an infinite system of nucleons
with a fixed ratio of neutron to proton numbers and no Coulomb
interaction. In general, the nuclear matter EOS, that is the
energy per particle, of asymmetric infinite nuclear matter
$\epsilon(n_{b},f_{a})$ \cite{2000:Chung} defined as
\begin{equation} \epsilon(n_{b},f_{a})=\frac{{\cal
{E}}}{n_{b}}\label{nuceos}\end{equation}
 is a function of two variables namely baryon number density $n_{b}$
and the relative neutron excess $f_{a}$ (the asymmetry parameter)
\begin{equation}
f_{a}=\frac{n_{n}-n_{p}}{n_{n}+n_{p}},\end{equation}
 where $n_{n}$ and $n_{p}$ denote the neutron and proton number
densities respectively. The sum $n_{n}+n_{p}=n_{b}$ stands for the
total baryon number density, ${\cal {E}}$ in equation (\ref{nuceos})
denotes the total energy of the nuclear system.\\
 The properties of asymmetric nuclear matter can be studied with
the use of the empirical parabolic approximation which allows one
to expand the energy per particle of asymmetric nuclear matter in
a Taylor series in $f_{a}$ \\
 \begin{equation}
\epsilon(n_{b},f_{a})=\epsilon(n_{b},0)+S_{2}(n_{b})f_{a}^{2}+S_{4}(n_{b})f_{a}^{4}+\ldots\label{eos0}\end{equation}
 The factor $f_{a}$ makes the quartic $S_{4}(n_{b})$ term contribution
negligible.\\
 Also the analysis performed with the use of realistic interactions
indicates the dominant role of the $S_{2}(n_{b})$ term not only in
the vicinity of the saturation point but even at higher densities
\cite{1991:bombaci}. The expansion given in equation (\ref{eos0})
enables the analysis of the function $\epsilon(n_{b},f_{a})$ in
terms of the energy of symmetric nuclear matter
$\epsilon(n_{b},0)$ and the symmetry energy $S_{2}(n_{b})$.
Subsequently expanding $\epsilon(n_{b},f_{a})$ around the
equilibrium density $n_{0}$ in a Taylor series in $n_{b}$, the
following expressions for the two successive terms
$\epsilon(n_{b},0)$ and $S_{2}(n_{b})$
can be obtained:\\
 \begin{equation}
\epsilon(n_{b},0)=\epsilon(n_{0})+\frac{1}{2}K_{v}x^{2}+\frac{1}{6}Q_{v}x^{3}+\ldots\label{en0}\end{equation}
 \begin{equation}
S_{2}(n_{b})=S_{2}(n_{0})+Lx+\frac{1}{2}K_{sym}x^{2}+\frac{1}{6}Q_{sym}x^{3}+\ldots\label{esym}\end{equation}
 where $x$ denotes dimensionless parameter that characterizes the
deviations of the density from its saturation value \begin{equation}
x=\frac{n_{b}-n_{0}}{3n_{0}}.\end{equation}
 Expressions (\ref{en0}) and (\ref{esym}) are parameterized by a
set of coefficients: $n_{0}$, $\epsilon(n_{0})$, $K_{v}$, $Q_{v}$,
$J$, $K_{sym}$, $L$ and $Q_{s}$ which determine the behavior of
the system near the saturation density. Particular coefficients are defined in Table \ref{eos} and
evaluated at the point $(n_{0},0)$. \\
This very point denotes the position of the state defined as the
equilibrium state of symmetric nuclear matter $\varepsilon(n_{0},0)$
with minimum energy per nucleon and is characterized by the condition
$\partial\varepsilon(n_{b},0)/\partial n_{b}=P(n_{0},0)=0$. Thus,
the linear term in the Taylor expansion (\ref{en0}) vanishes.\\
\begin{table}

\centering
\begin{tabular}{|c|c|}
\hline
Symmetric nuclear matter $\epsilon(n_{b},0)$  & Symmetry energy $S_{2}(n_{b})$
\tabularnewline
\hline
$\epsilon_{0}=\epsilon(n_{0},0)$  & $J=S_{2}(x=0)$
\tabularnewline
\hline
$K_{v}=9n_{0}^{2}\left(\frac{\partial^{2}{\epsilon(n_{b},0)}}{\partial{n_{b}^{2}}}\right)$ & $L=\left(\frac{\partial{S_{2}}}{\partial{x}}\right)$
\tabularnewline
\hline
$Q_{v}=27n_{0}^{3}\left(\frac{\partial^{3}{\epsilon(n_{b},0)}}{\partial{n_{b}^{3}}}\right)$ & $K_{sym}=(\frac{\partial^{2}{S}}{\partial{x^{2}}})$
\tabularnewline
\hline

\end{tabular}
\caption{Coefficients that parametrize the behavior of the symmetric nuclear
matter near saturation density.}
\label{eos}
\end{table}

 Gathering altogether the terms of the expansions in $n_{b}$ and
in $f_{a}$ the approximated form of the EOS can be written as
\cite{2000:Chung}\\
 \begin{equation}
\varepsilon(n_{b},f_{a})=\varepsilon(n_{0})+\frac{1}{18}(K_{0}+K_{sym}f_{a}^{2})\left(\frac{n_{b}-n_{0}}{n_{0}}\right)^{2}+\left[J+\frac{L}{3}\left(\frac{n_{b}-n_{0}}{n_{0}}\right)\right]f_{a}^{2}.\label{eq:eos0}\end{equation}
 Having obtained the EOS, each individual term that
enters the formula (\ref{eq:eos0}) can be calculated.\\
 According to the approximation presented by equation (\ref{eos0})
the symmetry energy can be calculated as the energy difference at
a given density between symmetric $(f_{a}=0)$ and pure neutron
matter $f_{a}=1$. The density dependance of the symmetry energy
around $n_{0}$ is determined by the parameters $L$ and $K_{sym}$.
Introducing the one-parameter fit to the low-density behavior of
the symmetry energy \[ E_{sym}(u)\approx J\, u^{\gamma}\]
 where $u=n/n_{0}$ and using this scaling properties the correlations
between the density dependance of the symmetry energy and the
neutron skin thickness can be estimated \cite{08-Li,
08:Piekarewicz, 09:Centeles}. This dependance also allows one to
determine the transition density $\rho_{t}$ between the crust and
the core of a neutron star and to express it through the
coefficients $K_{v}$ and $K_{sym}$ in the following way
\begin{equation}
u_{t}\sim\frac{2}{3}+\left(\frac{2}{3}\right)^{\gamma}\frac{K_{sym}}{2K_{v}}\end{equation}
 The constraints on the value of $\gamma_{t}$ obtained from the intermediate-energy
heavy-ion collisions provides a $\gamma$ value $\gamma\sim
0.69-1.05$. Calculations performed in this paper are based on the
standard TM1 parameter set \cite{94:Sugahara}. However, recent
experimental results strongly indicate lower value of the symmetry
energy coefficient and the compressibility coefficient of nuclear
matter \cite{08:Piekarewicz}. These lower values have been used to
construct a parameter set (denoted by RM) which when compared with
the TM1 one, differs in the value of the scalar meson field mass.
Also in the isovector sector the parameters $g_{\rho N}$ and
$\Lambda_{V}$ have been fitted  to reproduce the symmetry energy
coefficient at the value $J = 32$ MeV. The parameters and
saturation properties of symmetric nuclear matter are collected in
Table \ref{tab:TM1}.
\begin{table}
\centering\begin{tabular}{|c|c|c|c|c|}
\hline
 & $n_{0}$  & $a_{v}$  & $K_{v}$  & $Q_{v}$\tabularnewline
\hline
\hline
TM1  & $0.145$  & $-16.26$  & $281.16$  & $-258.295$\tabularnewline
\hline
FSUGold  & $0.148$  & $-16.27$  & $229.52$  & $-519.39$\tabularnewline
\hline
RM  & $0.148$  & $-16.0$  & $230.0$  & $-270.50$\tabularnewline
\hline
\end{tabular}

\caption{The parameters of the symmetric nuclear matter at saturation density.}
\label{tab:TM1}
\end{table}

\begin{table}
\centering\begin{tabular}{|c|c|c|c|c|c|}
\hline
 & $J$  & $L$  & $K_{sym}$  & $Q_{sym}$  & $K_{\tau}=K_{sym}-6\, L$\tabularnewline
\hline
\hline
TM1  & $36.89$  & $110.79$  & $33.52$  & $-74.28$  & $-631.26$\tabularnewline
\hline
FSUGold  & $32.6$  & $60.44$  & $-51.51$  & $414.51$  & $-414.15$\tabularnewline
\hline
RM  & $32.0$  & $75.04$  & $-57.22$  & $343.98$  & $-507.46$\tabularnewline
\hline
\end{tabular}

\caption{The parameters of the symmetric nuclear matter at saturation density.}

\end{table}

\begin{table}
\centering\begin{tabular}{|c|c|c|c|c|c|c|c|}
\hline
 & $\Lambda_{V}$  & $g_{\rho N}$  & $J\,(MeV)$  & $L\,(MeV)$  & $K_{sym}\,(MeV)$  & $K_{\tau}\,(MeV)$  & $\rho_{t}\,(fm^{-3})$\tabularnewline
\hline
\hline
TM1 (orig)  & $0.0$  & $9.2644$  & $36.89$  & $110.79$  & $33.52$  & $-631.25$  & $0.1026$\tabularnewline
\hline
TM1 (nonl)  & $0.0$  & $8.0642$  & $32.0$  & $96.12$  & $33.52$  & $-543.19$  & $0.1025$\tabularnewline
\hline
 & $0.008$  & $8.6567$  & $32.0$  & $85.13$  & $-37.98$  & $-548.77$  & $0.090$\tabularnewline
\hline
 & $0.01$  & $8.8264$  & $32.0$  & $82.38$  & $-50.93$  & $-545.23$  & $0.815$\tabularnewline
\hline
RM  & $0.0169$  & $\mbox{9.234}$  & $32.0$  & $75.04$  & $-57.22$  & $-507.46$  & $0.852$\tabularnewline
\hline
FSUGold  & $0.03$  & $11.767$  & $32.59$  & $60.44$  & $-51.51$  & $-414.15$  & $0.857$\tabularnewline
\hline
\end{tabular}

\caption{Saturation Properties of nuclear matter at saturation density obtained
for nonlinear models}

\end{table}

\begin{figure}
\centering \includegraphics[clip,width=9cm]{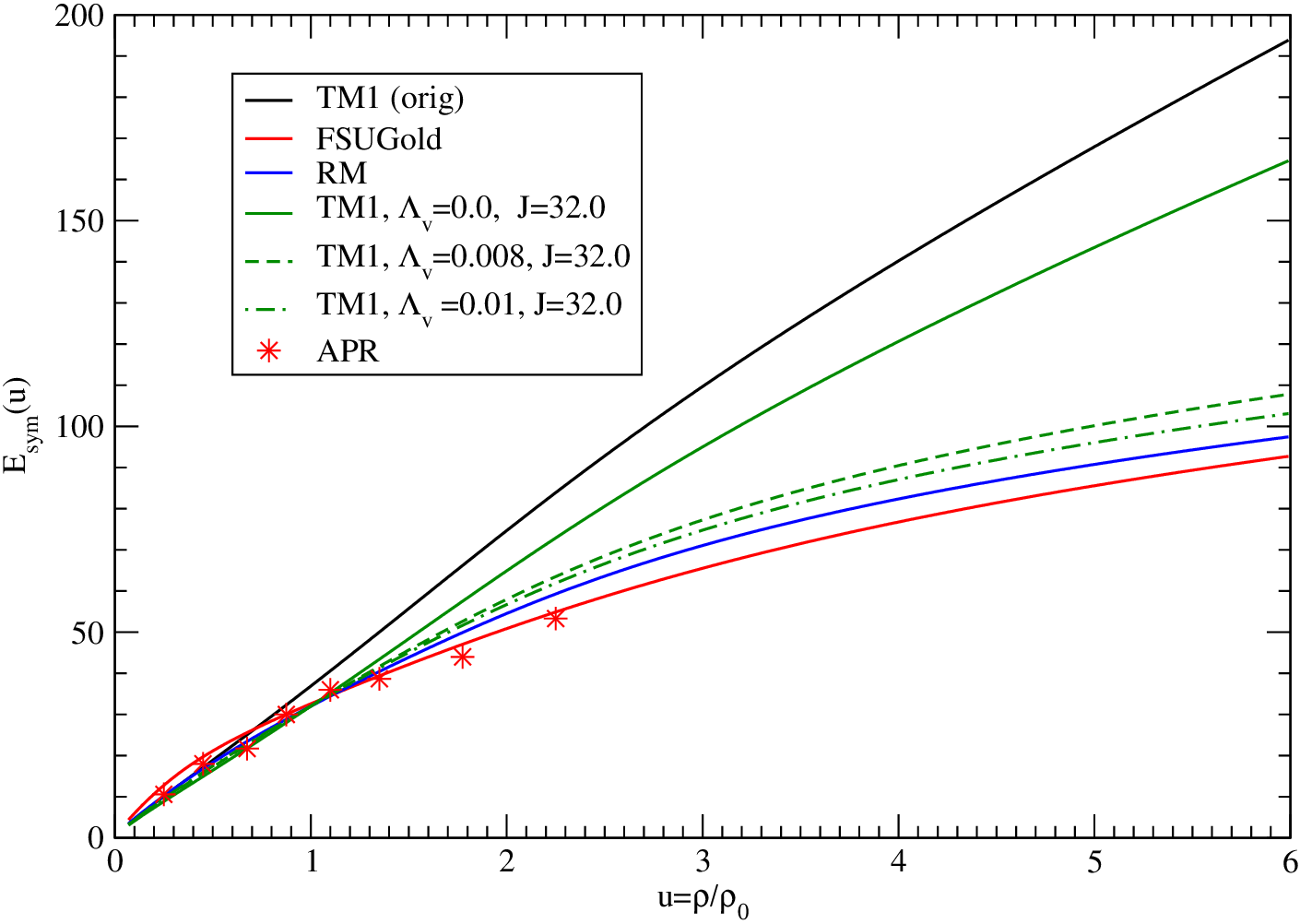} \label{ES}

\caption{ The density dependence of the symmetry energy for
nonlinear models. For comparison the results obtained for the
ordinary TM1 parameter set has been included. }

\end{figure}

The inclusion of the  mixed nonlinear isoscalar-isovector coupling
$\Lambda_V$
  provides the
additional possibility of modifying the high density components of
the symmetry energy and  requires the adjustment of the $g_{\rho
N}$ coupling constant to keep the same value of the symmetry
energy at saturation. The remaining ground state properties are
left unchanged. With these  additional terms the  expression for
the symmetry energy coefficient $E_{sym}(n_0)$ is now given by the
equation
\begin{eqnarray}
E_{sym}(n_0)&=&\frac{1}{8}\frac{n_0}{m_{\rho}^2/g_{\rho
N}^2+2\Lambda_V(g_{\rho N}g_{\omega
N})^2w_0^2}\nonumber \\
&+&\frac{k_F^2}{6\sqrt{k_F^2+m_{0}^2}},
\end{eqnarray}
where where $k_0$ and $m_0$ are the Fermi momentum and nucleon
effective mass of symmetric nuclear matter at saturation. The
first term in this equation coming from the explicit coupling
between the nucleon isospin and the $\rho$ meson whereas the
second quantity is the relativistic kinetic energy contribution.
The influence of the nonlinear couplings can also be considered in
terms of effective $\omega$ and $\rho$ meson masses
\cite{2001:Manka} which can be defined by the following relation
\begin{equation}
m_{eff,\omega}^2=m_{\omega}^2+2\Lambda_V(g_{\rho N}g_{\omega
N})^2r_0^2,
\end{equation}
\begin{equation}
m_{eff,\rho}^2=m_{\rho}^2+2\Lambda_V(g_{\rho N}g_{\omega
N})^2w_0^2.
\end{equation}
In this interpretation this is the $\rho$ meson mass modification
that influences the density dependence of the symmetry energy. The
obtained form of the symmetry energy for the considered parameter
sets are presented in Fig. \ref{ES}. In general the nonlinearities
soften the density dependance of the symmetry energy. For
comparison the results of Akmal et al. have been included
\cite{98:Akmal}

Vector mesons-hyperon coupling constants are taken from the quark
model and they are given by relations (\ref{veccoup}). Whereas in
the scalar sector the scalar coupling of the $\Lambda$ and $\Xi$
hyperons requires constraining in order to reproduce the estimated
values of the potential felt by a single $\Lambda$ and a single
$\Xi$ in the saturated nuclear matter. The analysis of the
experimental data concerning the binding energies of $\Lambda$'s
bound in single particle orbitals in hypernuclei over an extensive
range of mass numbers makes it possible to determine the potential
depth of a single $\Lambda$ in nuclear matter at the value of
\begin{equation} U_{\Lambda}^{(N)}\simeq 27-30\, MeV\end{equation}
 which corresponds to $\sim$ 1/2 of the nucleon well depth $U_{N}^{(N)}$.
There is still considerable uncertainty about the experimental
status of $\Sigma$-nucleus potential. The calculations of $\Sigma$
hypernuclei have been based on analysis of $\Sigma^{-}$ atomic
data. Phenomenological analysis of level shifts and widths in
$\Sigma^{-}$ atoms made by Batty et al. \cite{ksi, 07:Friedman}
indicates that the $\Sigma$ potential is attractive only at the
nuclear surface, becoming repulsive for increasing density. The
small attractive component of this potential is not sufficient to
form bound $\Sigma$-hypernuclei. Also according to recent
experimental data it has been established that the $\Sigma$
nuclear
interaction is strongly repulsive.\\
 The following values of the potentials have been used \cite{02:Schaffner} \begin{equation}
U_{\Lambda}^{(N)}=-28MeV,\,\, U_{\Sigma}^{(N)}+30MeV,\,\,
U_{\Xi}^{(N)}=-18MeV\end{equation}
 for the determination of the $g_{\sigma\Lambda}$, $g_{\sigma\Sigma}$
and $g_{\sigma\Lambda}$ coupling constants. In order to properly
describe hyperon rich neutron star matter, the knowledge of the
hyperon-hyperon interaction is indispensable. Data on
$\Lambda\Lambda$ hypernuclei are scarce. Observation of
double-strange hypernuclei $\Lambda\Lambda$ provide information
about the $\Lambda-\Lambda$ interaction. Several events have been
identified which indicate an attractive $\Lambda\Lambda$
interaction. The analysis of the data allows one to estimate the
binding energies of ${}_{\Lambda\Lambda}^{6}$He,
${}_{\Lambda\Lambda}^{10}$Be and ${}_{\Lambda\Lambda}^{13}$B. The
interaction between other type of hyperons are not known
experimentally \cite{96:Schaffner, 08:Bielich}. The hyperon
couplings to strange meson $\sigma^{\ast}$ have been obtained from
the following relations \begin{equation} U_{\Xi}^{(\Xi)}\simeq
U_{\Lambda}^{(\Xi)}\simeq2U_{\Xi}^{(\Lambda)}\simeq2U_{\Lambda}^{(\Lambda)}.\end{equation}
 In summery the coupling of hyperons to the strange meson $\sigma^{\ast}$
has been limited by the estimated value of hyperon potential
depths in hyperon matter this has also direct consequences for
neutron star parameters. Recent experimental data \cite{Takahashi}
indicate a much weaker strength of hyperon-hyperon interaction.
The currently obtained value of the $U_{\Lambda}^{(\Lambda)}$
potential at the level of 5 MeV permits the existence of the
additional parameter set which reproduces this weaker
$\Lambda\Lambda$ interaction. The parametrisation considered in
this paper includes for comparison the strong and weak
hyperon-hyperon couplings. The strong $Y-Y$ interaction is related
to the value of the potential $U_{\Lambda}^{(\Lambda)}=-20$ MeV,
whereas the weak one corresponds to $U_{\Lambda}^{(\Lambda)}=-5$
MeV \cite{Takahashi},
\cite{Song}.\\
 The inclusion of additional parameters in the isovector meson
sector requires the adjustment of the $g_{\rho N}$ parameters. The
new values of parameters are collected in Table \ref{tab:gro}.

\begin{table}
\centering\begin{tabular}{|l|l|l|l|l|}
\hline
Y--Y interaction  & $g_{\sigma\Lambda}$  & $g_{\sigma\Xi}$  & $g_{\sigma^{\ast}\Lambda}$  & $g_{\sigma^{\ast}\Xi}$\tabularnewline
\hline
weak  & 6.17  & 2.202  & 5.41 & 11.516 \tabularnewline
\hline
strong  & 6.17  & 3.202  & 7.018  & 12.6 \tabularnewline
\hline
\end{tabular}

\caption{Strange scalar sector parameters}
\label{tab:sscalar}
\end{table}

\begin{table}

\centering\begin{tabular}{|l|l|l|l|}
\hline
&$\Lambda_{V}=0$  & $\Lambda_{V}=0.008$  & $\Lambda_{V}=0.01$\tabularnewline
\hline
$g_{\rho N}$ &9.2644  & 10.207  & 10.4828\tabularnewline
\hline
\end{tabular}

\caption{The $g_{\rho N}$ parameters for different values of the
parameter $\Lambda_{V}$}

\label{tab:gro}
\end{table}

\section{Results}

Having obtained the EOS that relates the energy density and
pressure the corresponding solution of the
Tolman-Oppenheimer-Volkoff (TOV) equations can be found and
estimation of neutron star masses and radii become possible. The
EOS and especially its high density limit has inevitable
consequences for neutron star parameters. This manifests itself in
a deep sensitivity of neutron star masses and radii to the
stiffness of the EOS and allows one to study the influence of
nonlinear vector meson interaction terms on neutron star
properties.
\begin{figure}
\centering \includegraphics[clip,width=9cm]{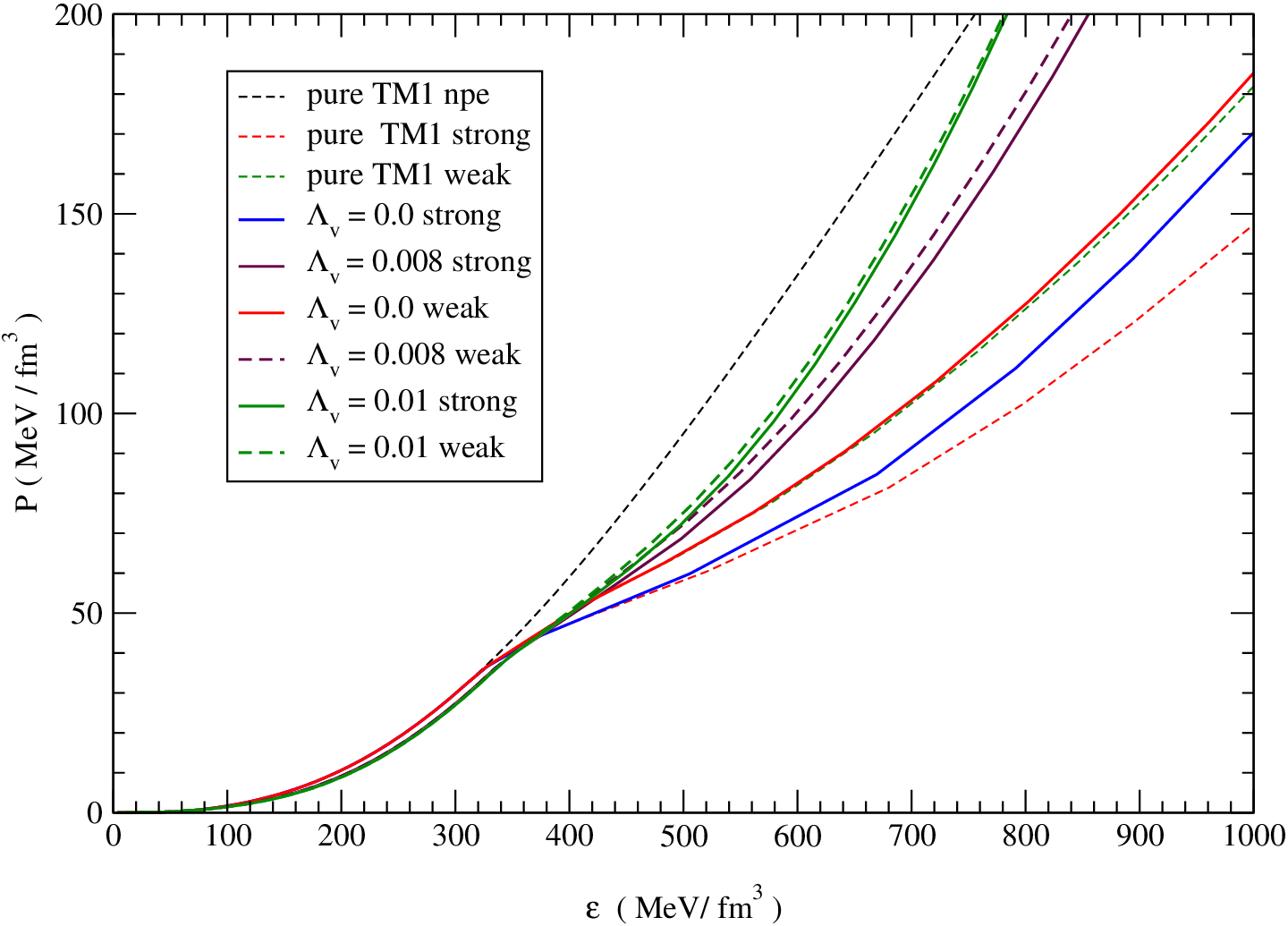} 

\caption{The EOS obtained for the nonlinear models. The limiting
cases of the EOS indicate that the stiffest one has been obtained
for the non-strange matter, whereas the softest represents the
case of the standard TM1 model extended by the inclusion of
strange mesons which have been introduced in a minimal fashion.
The remaining EOS for the nonlinear models with different values
of the parameter $\Lambda_{V}$ are located between these two
curves.} \label{fig:eos}
\end{figure}

\begin{figure}
\centering \includegraphics[width=9cm]{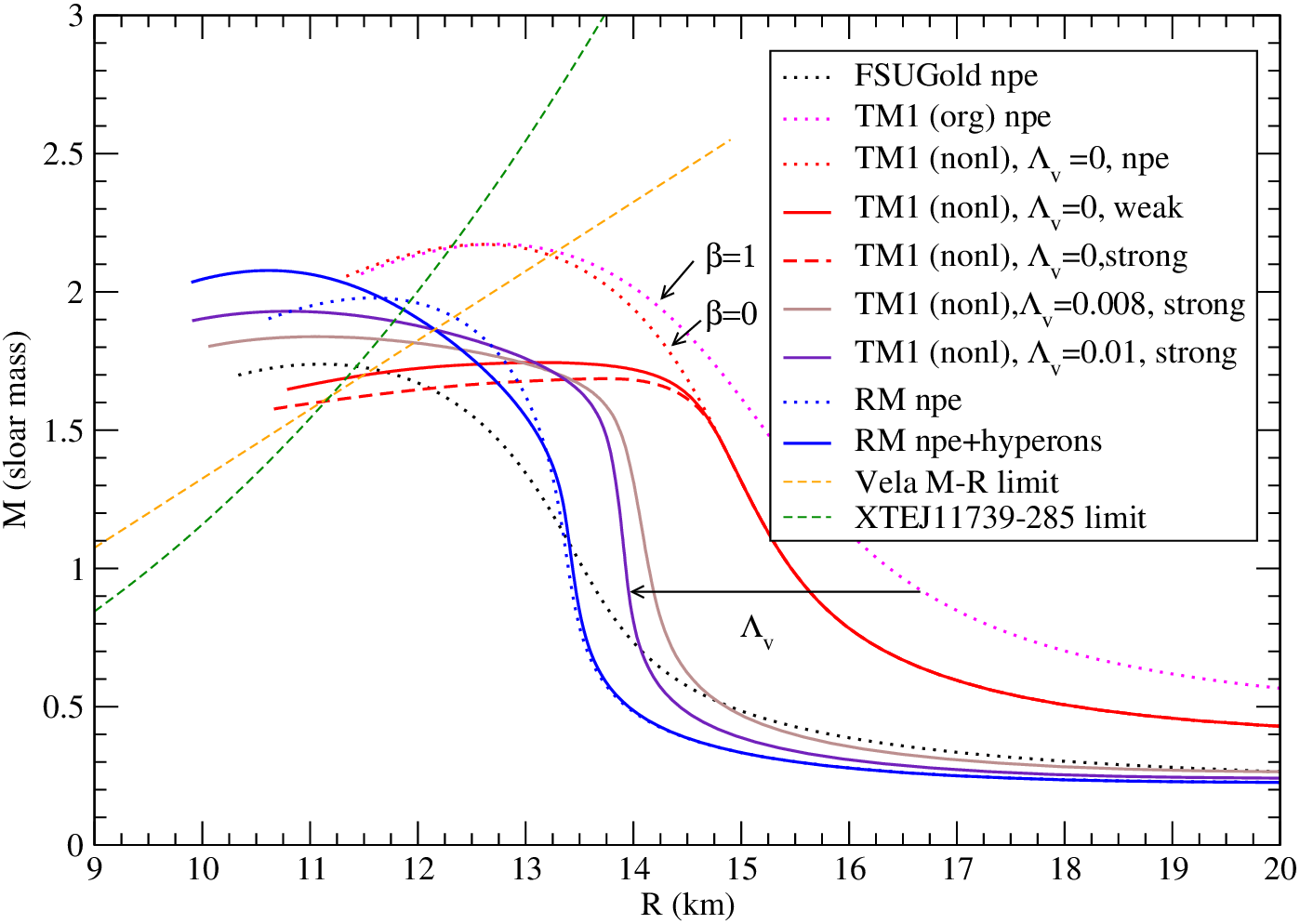}

\caption{The mass-radius relations obtained for the chosen models.}
\label{fig:RM}
\end{figure}

The integration of the TOV equations with a specific equation of
state leads to the mass-radius relation and allows one to
determine the value of the maximum mass which in a sense can give
a measure of the impact of particular nonlinear couplings between
vector mesons. In Fig. \ref{fig:eos} the equations of state
obtained for different cases of nonlinear potentials presented in
this paper have been  shown. Extreme, dashed curves represent
results obtained for the standard TM1 parameterisation, for the
non-strange and strangeness rich matter respectively. The case
when the matter includes only nucleons and leptons gives the
stiffest EOS whereas the directly opposed result namely the
softest EOS can be obtained for the standard TM1 parameterisation
extended by the inclusion of hyperons. The vector meson sector in
this case comprises the quartic vector meson $\omega_{\mu}$
self-interaction term supplemented by the linear term for the
hidden-strange meson $\phi_{\mu}$. The influence of the strength
of hyperon-hyperon interaction is also illustrated by comparing
equations of state calculated for the weak and strong $Y-Y$
interaction. Other equations of state presented in this figure aim
to provide the analysis of the influence of additional vector
meson nonlinear interaction. Relating this problem to the
introduced scheme for nonlinear potentials, the consider cases
$\Lambda_{V}=0.008$ and $\Lambda_{V}=0.01$ are taking into
account. The obtained results indicate for the strong tendency for
stiffening of the EOS for the increasing value of the parameter
$\Lambda_{V}$, approaching the limiting case for
$\Lambda_{V}=0.01$. Moreover, there exists evidence for
diminishing the differences between the weak and strong $Y-Y$
interaction for increasing value of the parameter $\Lambda_{V}$.
Calculations performed for the value of the parameter
$\Lambda_{V}>0.01$ lead to acausal behavior of the equation of
state at high density.
\\
 The inclusion of nonlinear vector meson interactions has profound
consequences for the structure of neutron stars and this can be
deduced from Fig. \ref{fig:RM} where the mass-radius relations for
the obtained equations of state have been shown. Dotted curves
depicts the results calculated for  non-strange matter. Arrows
marked by $\beta = 0$ and $\beta =1$ denotes the solutions
obtained for ordinary TM1 parameter set and for the TM1
supplemented by the nonlinear $\omega-\rho$ coupling between the
isoscalar and the isovector mesons. In the  case of strangeness
rich matter the results for different values of the parameter
$\Lambda_{V}$ has been included. The arrow marked $\Lambda_{V}$
shows the influence of the nonlinear terms on neutron star
parameters especially for radii.  For comparison the mass-radius
relation for the FSU Gold parameter set has been given. This
figure depicts also constraints obtained from Vela and XTEJ
11739-285 data \cite{09:Li}.

\begin{figure}
\centering \includegraphics[width=9cm]{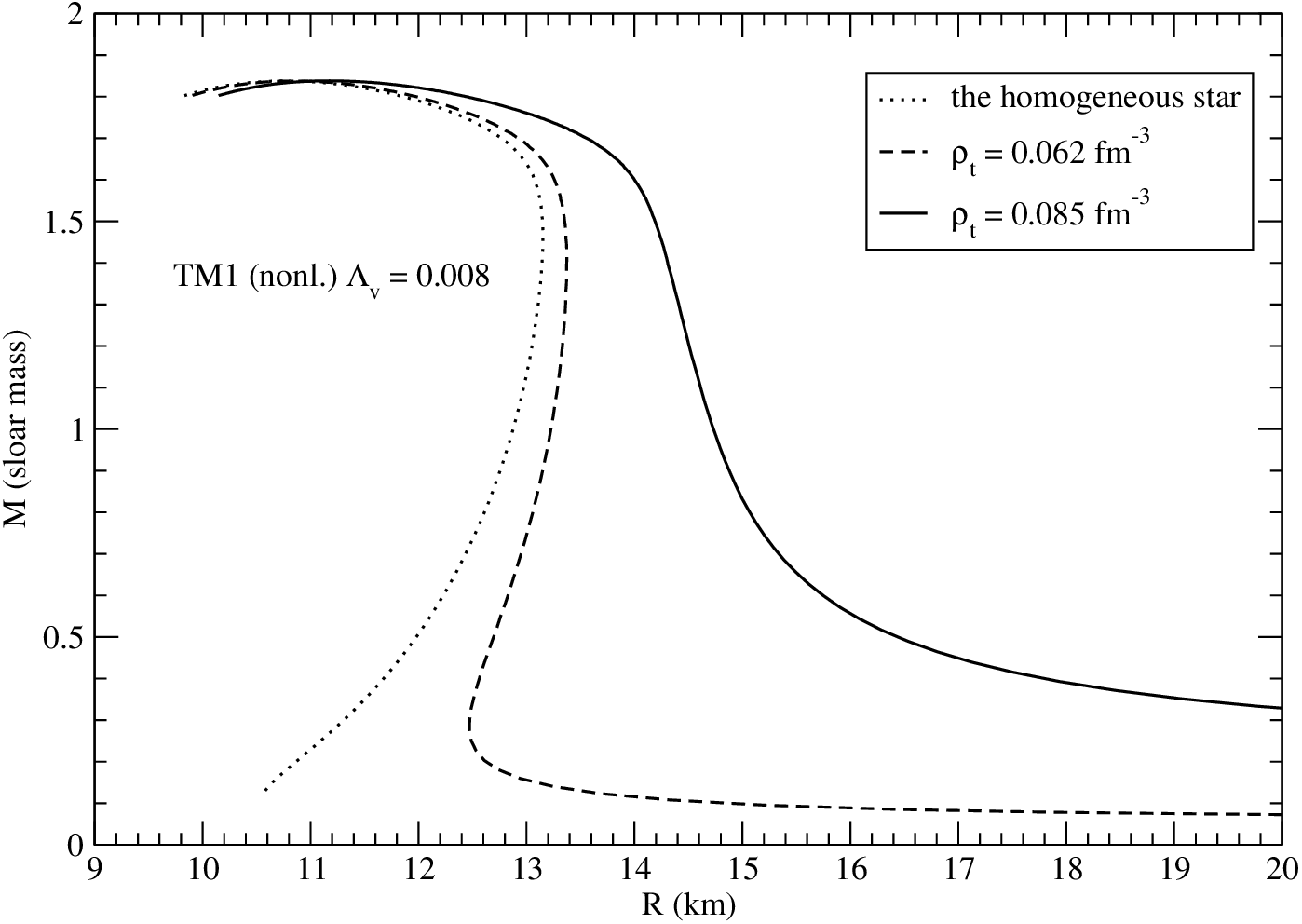}

\caption{The mass-radius relation for the chosen models. It includes for comparison
homogenous model without outer envelope.}
\label{RM008}
\end{figure}

Fig \ref{RM008} depicts models obtained for the value of
$\Lambda_{V}=0.008$ with and without the nonhomogenous inner crust
phase. This figure  compares the solutions obtained for different
structure of the outer layer of the neutron star. The presented
results includes the homogenous neutron star model without the
crust which is represented by dotted curve whereas dashed and
solid lines depicts the mass-radius relations for the crust
without and with the nonhomogenous inner crust, respectively. For
both cases the critical density $\rho_{t}$ is given. A comparison
of Fig. \ref{fig:RM} and Fig. \ref{RM008} leads to the conclusion
that results obtained for higher value of the parameter
$\Lambda_{V}$ gives lower value of the critical density $\rho_{t}$
and as a consequence diminishes the nonhomogenous inner crust.

The values of the maximum masses and the corresponding values of
radii for the strangeness rich matter have been collected in Table
\ref{tab:RM}. Results have been calculated for selected values of
the parameter $\Lambda_{V}$.

\begin{table}

\label{tab:RM}

\centering\begin{tabular}{|l|l|l|l|}
\hline
$\Lambda_{V}$  & $\rho_{max}(gcm^{-3})$  & $M_{\max}(M_{\odot})$  & $R(M_{\max})$\tabularnewline
\hline
$\Lambda_{V}=0$  & 14.4  & 1.62  & 12.92\tabularnewline
\hline
$\Lambda_{V}=0.008$  & 27.4 & 1.95 & 10.53\tabularnewline
\hline
$\Lambda_{V}=0.01$  & 28.7  & 2.161  & 10.06\tabularnewline
\hline
\end{tabular}
\caption{The value of the maximum mass configurations  and the corresponding radii for different values of the parameter $\Lambda_{V}$ }
\end{table}

\begin{figure}
\begin{centering}
\centering\includegraphics[clip,width=8cm]{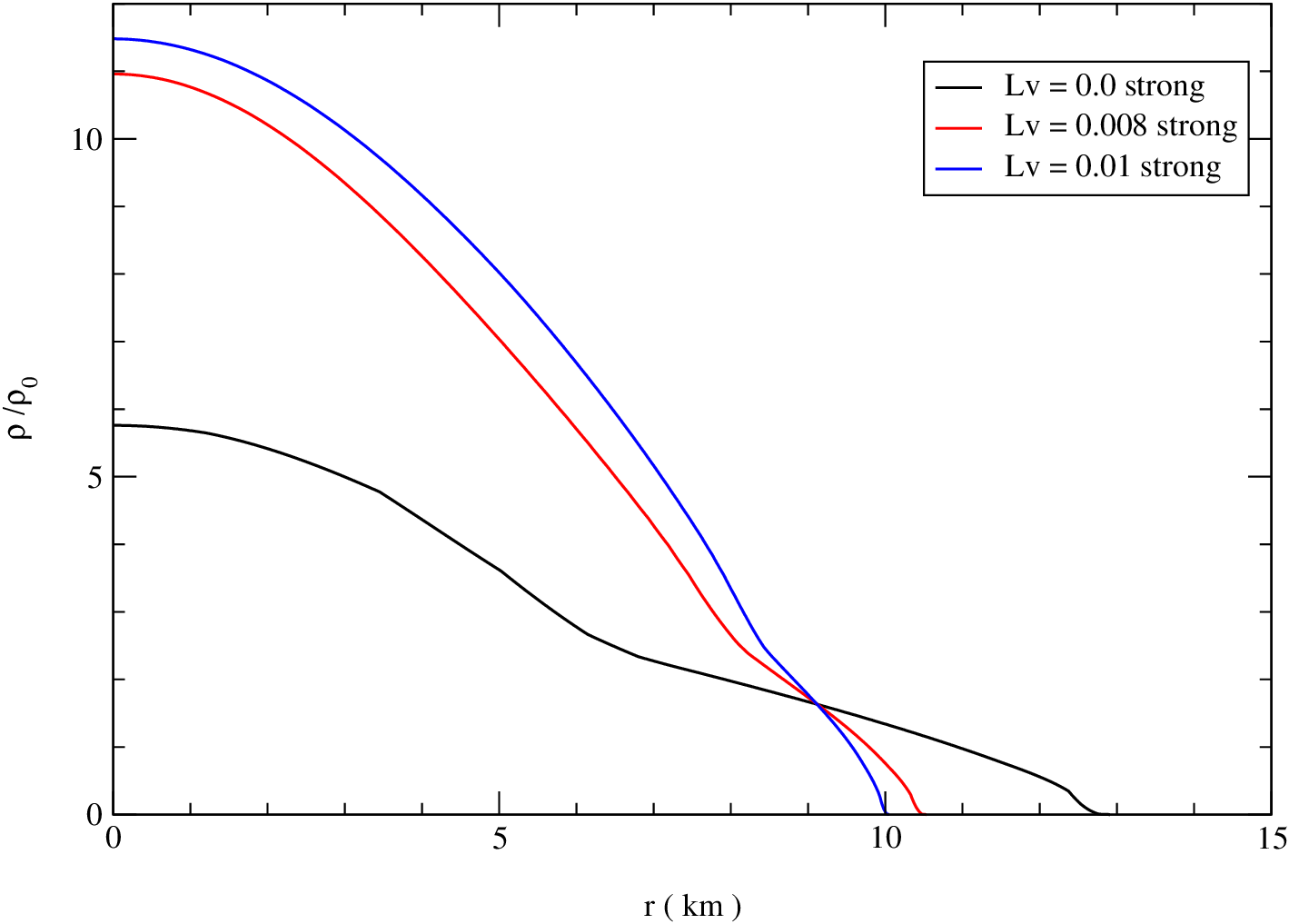}

\par\end{centering}

\caption{The density profiles in the models representing the
maximum mass configurations.}
 \label{fig:rhoR}
\end{figure}

The maximum mass increases with increasing value of the parameter
$\Lambda_{V}$, given in the results neutron star models with
masses exceeding $2M_{\odot}$ and with reduced radii. Thus, one
can expect that solutions with nonlinear vector meson couplings
lead to neutron star models with substantially greater density.
The mass-radius diagrams have been obtained for variety of models
presented in this paper. It includes the cases for non-strange and
strange matter. Firstly solution for the pure TM1 model only with
quartic $\omega$ term is compared with the enlarged nonlinear TM1
one which additionally comprises quartic $\rho$ meson term. From
these two mass-radius relations it is evident that the presence of
the quartic $\rho$ meson term influences mainly neutron star
radii. Changing the value of the parameter $\Lambda_{V}$ solutions
with substantially reduced value of the transition density is
obtained. This points to the conclusion that the crust-core
boundary moves to lower density region leading to the models with
reduced value of non-homogenous phase. This is confirmed in Fig.
\ref{fig:rhoR} which shows the density profile of neutron star
matter for maximum mass configuration. Models with nonlinear
vector meson couplings give as a result neutron stars with
densities much more higher than that obtained for the case
$\Lambda_{V}=0$.

The analysis of the density profiles indicates for the existence of
different regions in neutron star interiors. These regions corresponds
to the envelope and the core. In the core there is distinct part,
with substantially increased density. This is connected with the appearance
of strange matter. In the case of nonlinear models the inner core
with nonzero strangeness spread through-out almost the whole star
leading to very uniform neutron star model with considerably reduced
outer parts.\\
 The appearance of hyperons follows from the chemical potential
relations. It has been shown that the composition of hyperon star
matter as well as the threshold density for hyperons, is altered when
the strength of the hyperon-hyperon interaction is changed. %
\begin{figure}
\begin{centering}
\centering\includegraphics[clip,width=8cm]{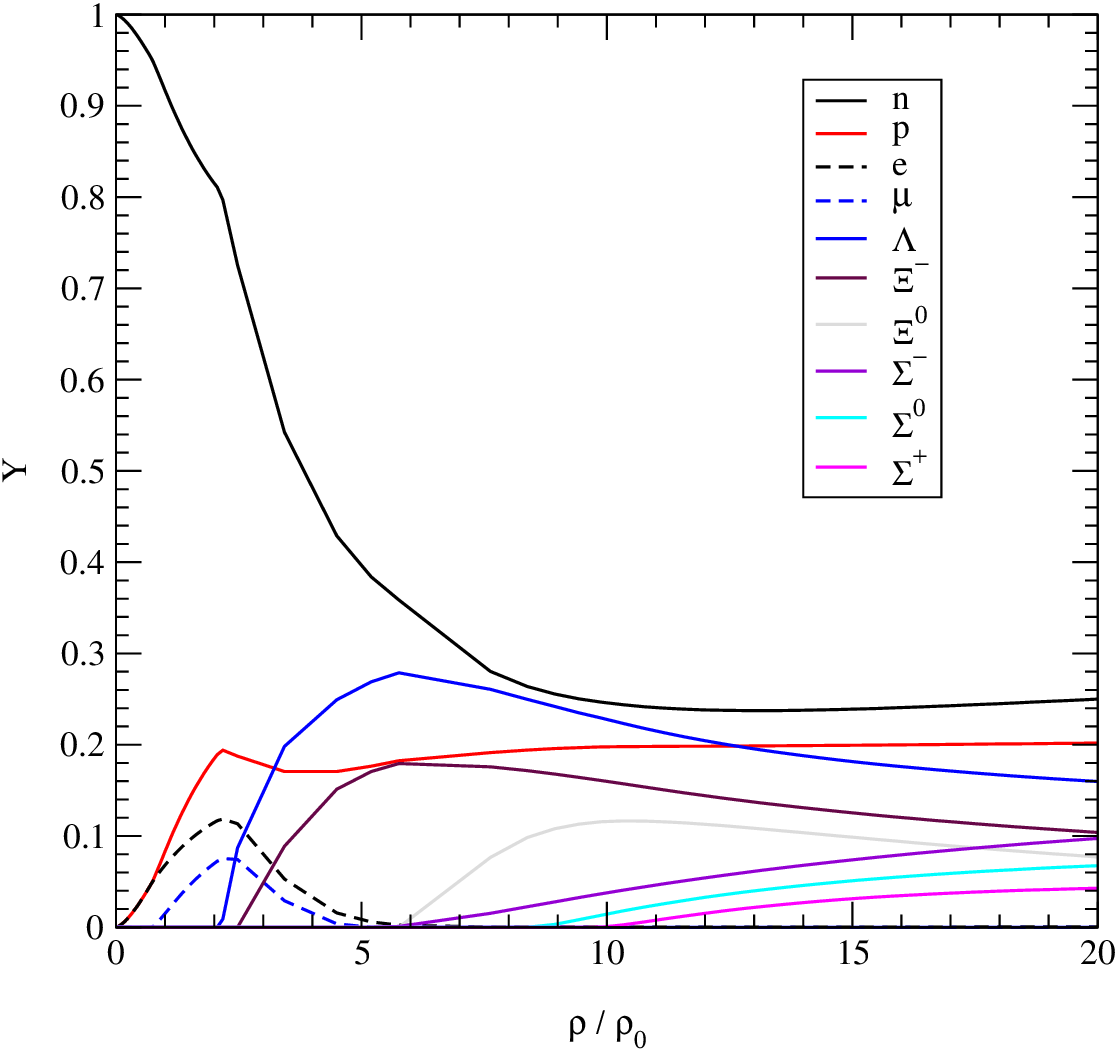}
\par\end{centering}

\caption{Relative concentrations of baryons for the model with $\Lambda_{V}=0$,
for the strong $Y-Y$ interaction.}

\label{YBstrongLin}
\end{figure}

However, the inclusion of nonlinear vector meson interactions also
significantly modifies the chemical composition of the star. In order
to get more complete understanding of the influence of nonlinear vector
meson coupling on particular baryon and lepton concentration it is
interesting to analyze this issue from different perspectives.%
\begin{figure}
\begin{centering}
\centering\includegraphics[clip,width=8cm]{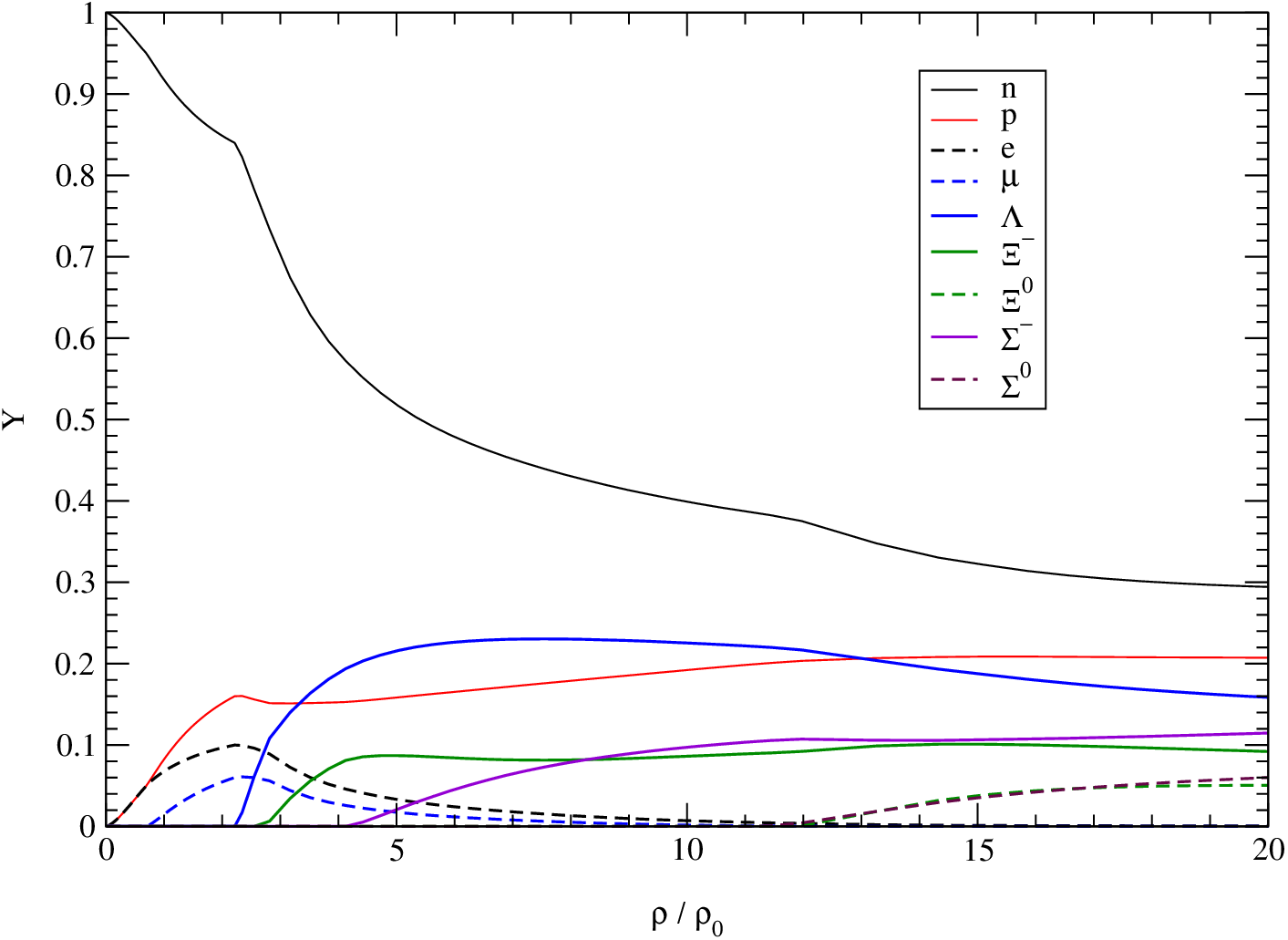}
\par\end{centering}

\caption{Relative concentrations of baryons for the model with nonlinear vector
coupling for $\Lambda_{V}=0.008$, for the strong $Y-Y$ interaction.}

\label{YBstrongNl}
\end{figure}

\begin{figure}
\begin{centering}
\centering\includegraphics[clip,width=8cm]{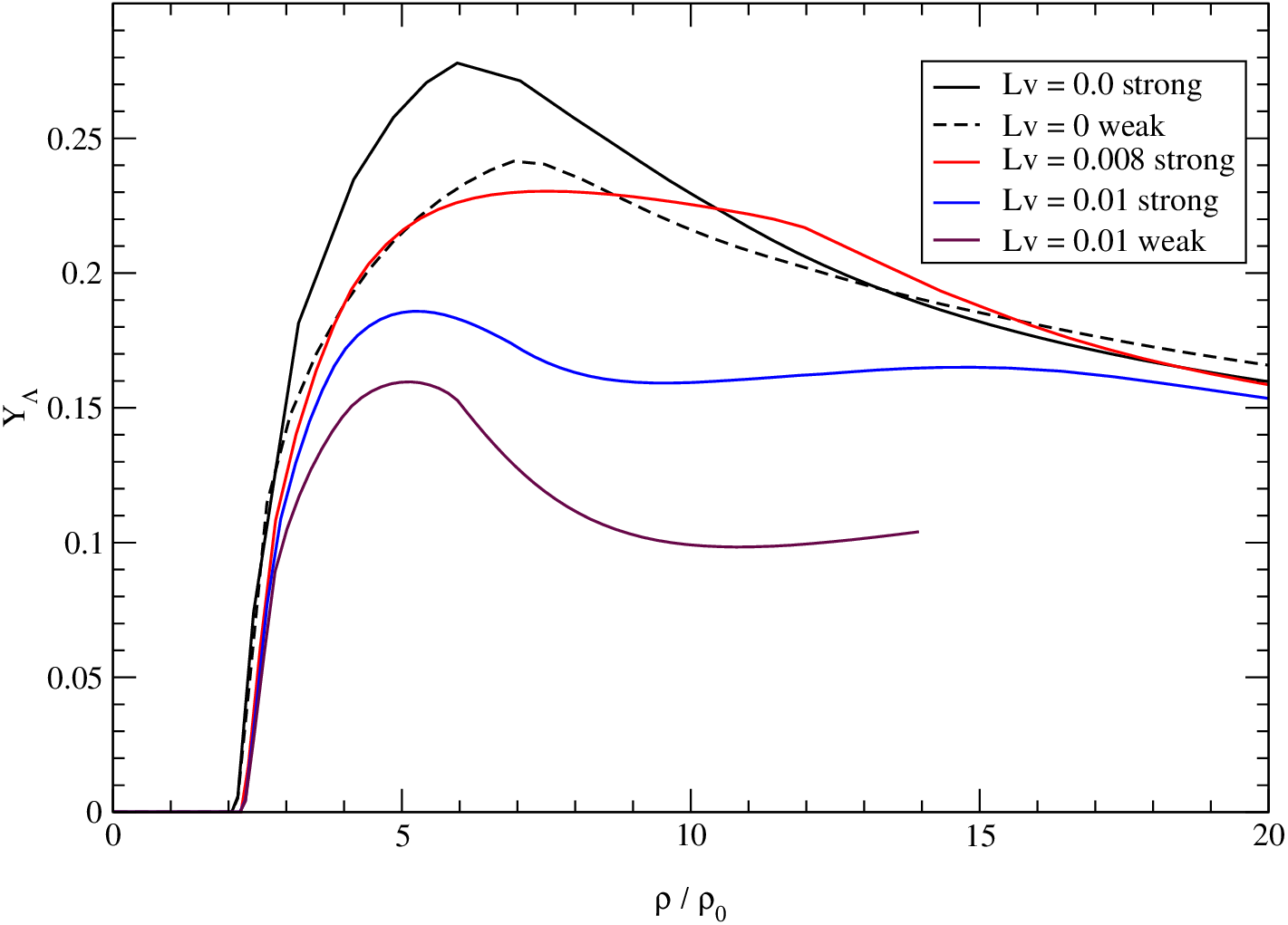}
\par\end{centering}

\caption{Concentrations of $\Lambda$ hyperons for the cases discussed in the
text, for different values of the parameter $\Lambda_{V}$. For comparison
the results obtained for the weak hyperon-hyperon interaction is also
included. The weak models are constructed for $\Lambda_{V}=0$ and
$\Lambda_{V}=0.008$.}

\label{YL}
\end{figure}

\begin{figure}
\begin{centering}
\centering\includegraphics[clip,width=8cm]{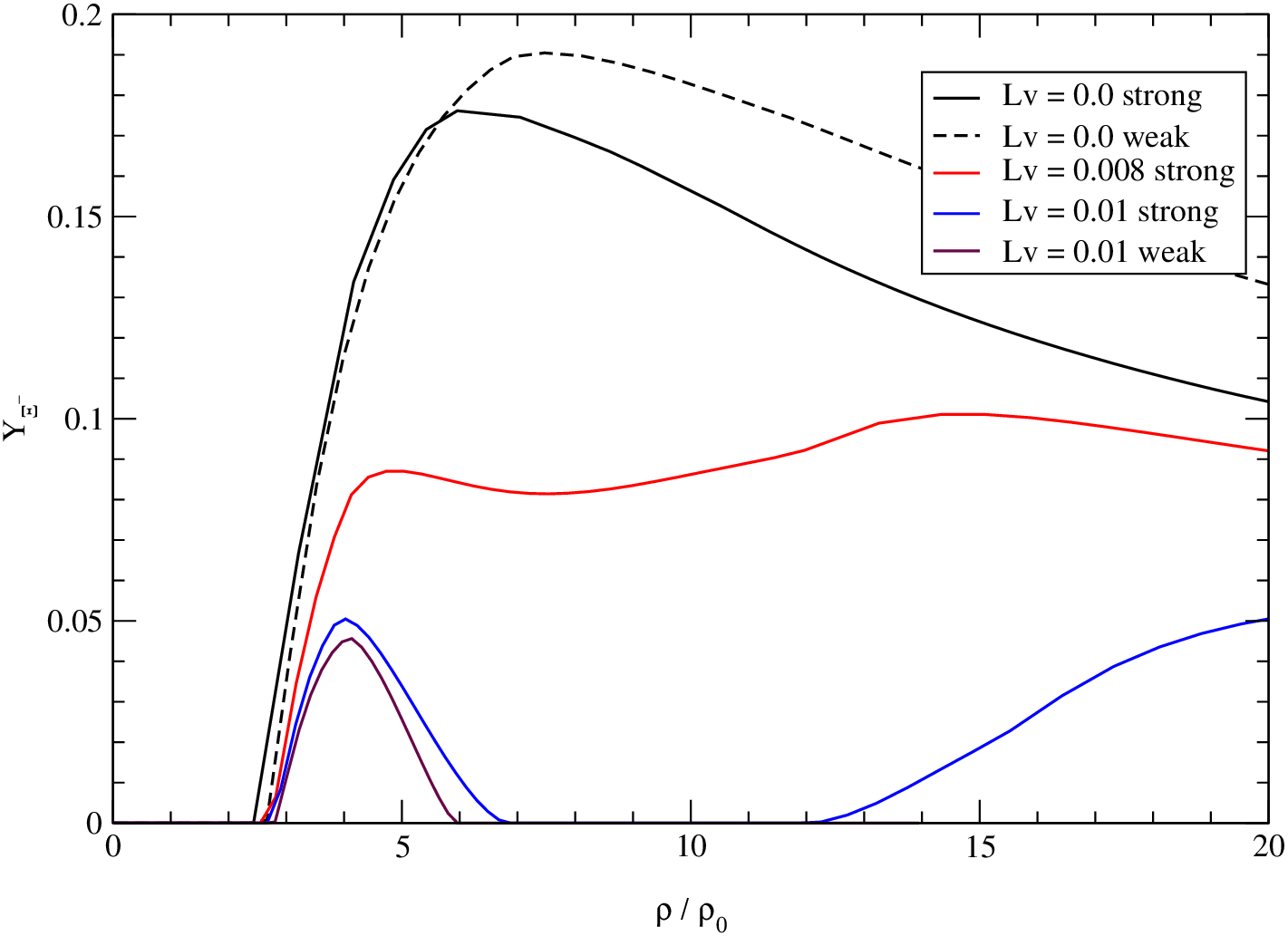}
\par\end{centering}

\caption{Concentrations of $\Xi^{-}$ hyperons for the cases discussed in the
text, for different values of the parameter $\Lambda_{V}$. For comparison
the results obtained for the weak hyperon-hyperon interaction is also
included. The weak models are constructed for $\Lambda_{V}=0$ and
$\Lambda_{V}=0.008$.}

\label{YX}
\end{figure}

Fig. \ref{YBstrongLin} and \ref{YBstrongNl} show relative fractions
of particular baryon species $YB_{i}$ ($i=n,p,\Lambda,\Xi^{-},\Xi^{0},\Sigma^{-},\Sigma^{+}$
and $\Sigma^{0}$) as a function of baryon number density $n_{b}$
for the strong value of $Y-Y$ interaction. Fig. \ref{YBstrongLin}
is constructed for the model without nonlinear vector meson coupling
whereas Fig. \ref{YBstrongNl} for the nonlinear model with $\Lambda_{V}=0.008$.
All calculations have been done under the assumption that the repulsive
$\Sigma$ interaction shifts the onset points of $\Sigma$ hyperons
to very high densities.\\
 $\Lambda$ is the first strange baryon that emerges in hyperon
star matter, it is followed by $\Xi^{-}$ and $\Xi^{0}$ and $\Sigma$
hyperons. These figures show differences in baryon fractions but these
differences are more clearly visible in Fig. \ref{YL} and \ref{YX}
presenting $\Lambda$ and $\Xi$ hyperon concentrations for different
models. %

\begin{figure}
\begin{centering}
\centering\includegraphics[clip,width=8cm]{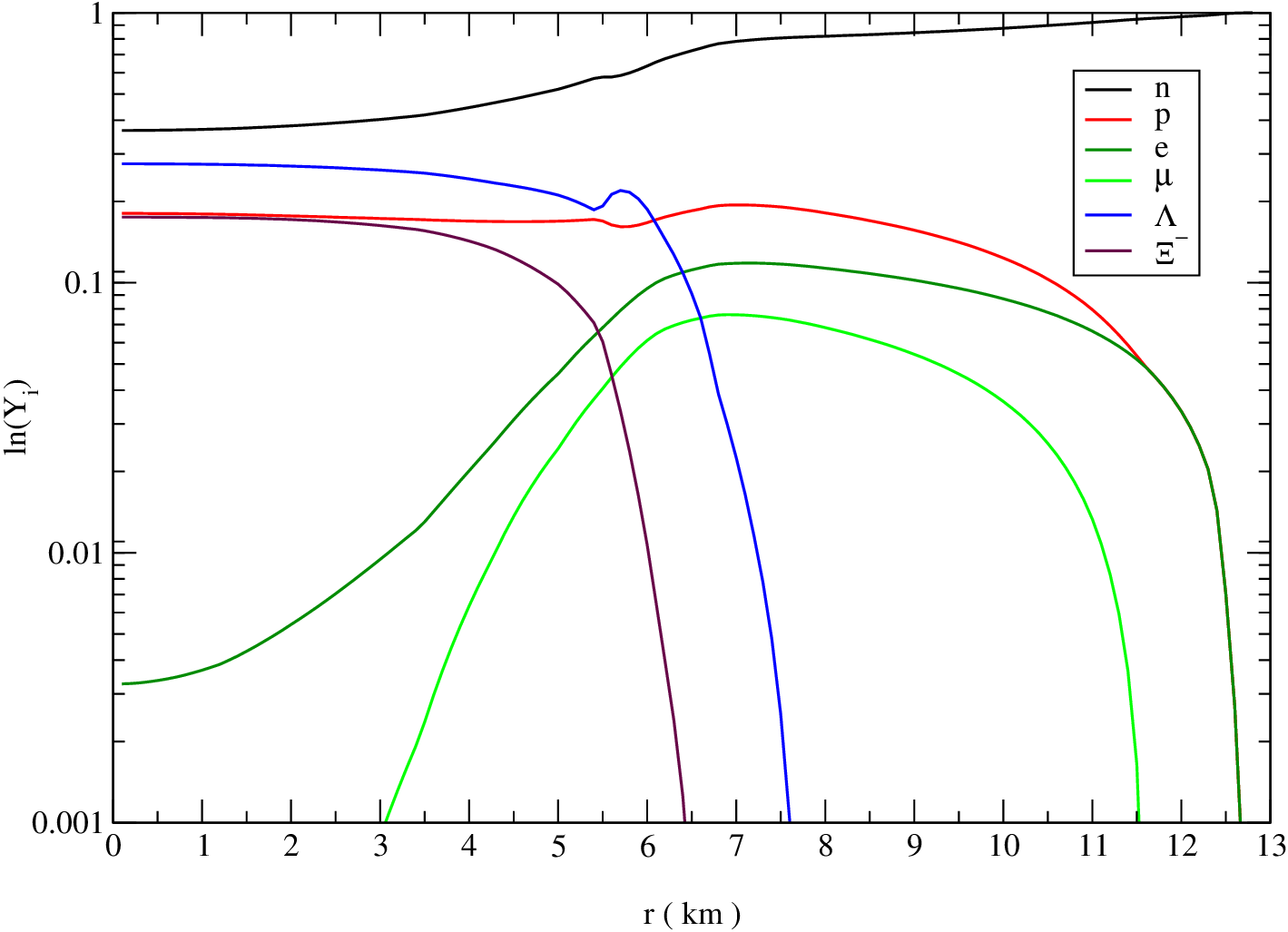}
\par\end{centering}

\caption{Baryon concentrations as a function of stellar radius for $\Lambda_{V}=0$.}

\label{YR0}
\end{figure}

\begin{figure}
\begin{centering}
\centering\includegraphics[clip,width=8cm]{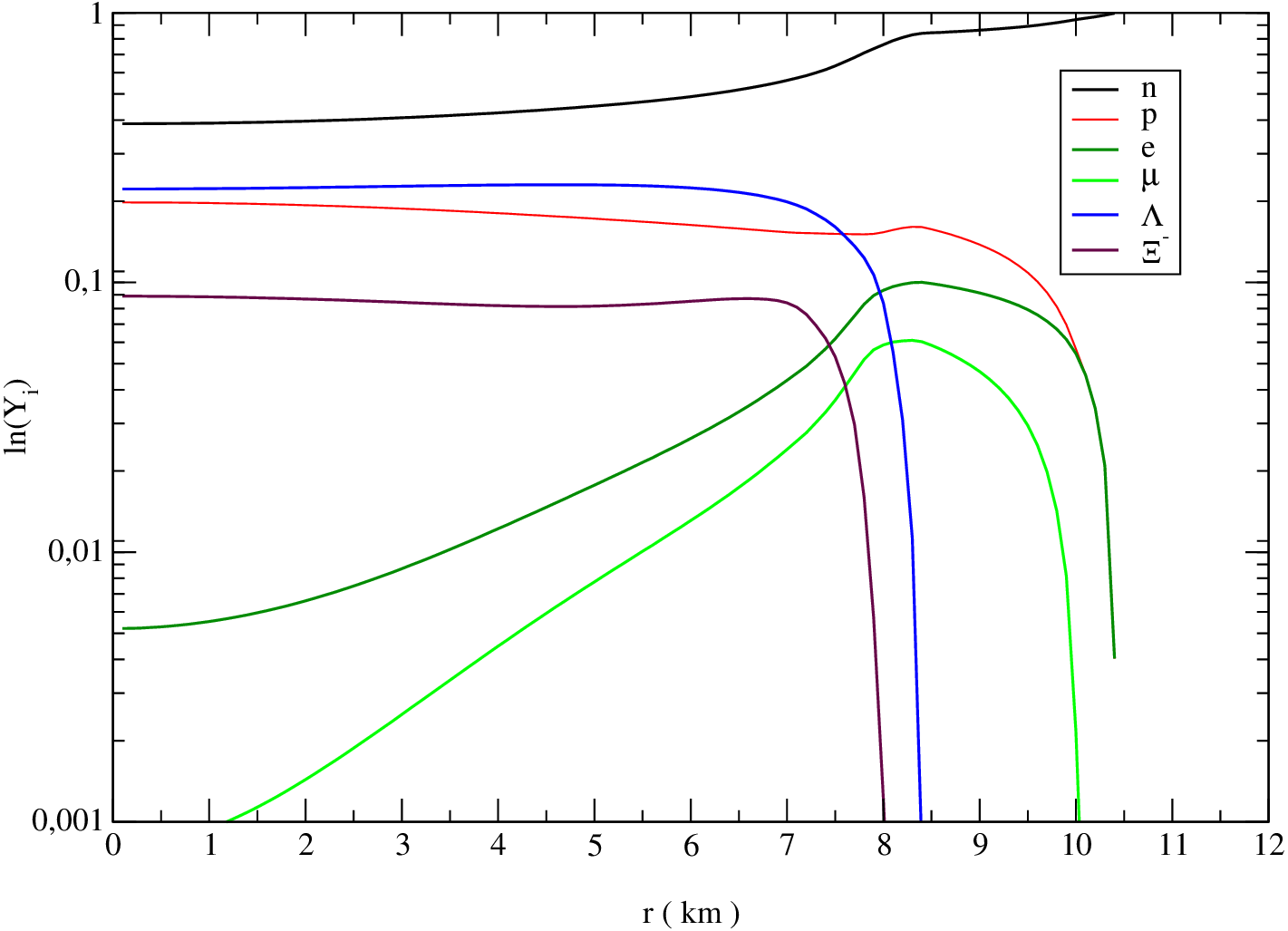}
\par\end{centering}

\caption{Baryon concentrations as a function of stellar radius for $\Lambda_{V}=0.008$}

\label{YR008}
\end{figure}

Fig. \ref{YL} and \ref{YX} depict the analysis done for several
models focusing on comparing the results for different values of
the parameter $\Lambda_{V}$. Generally the appearance of nonlinear
vector meson interaction terms lowers the concentration of
hyperons in neutron star matter. Thus there is a substantial
reduction of the $\Lambda$ and $\Xi^{-}$ hyperon populations for
the increasing value of $\Lambda_{V}$. In the case of $\Xi$
hyperon there is a density range for which the population of $\Xi$
hyperons vanishes. This results from  the chemical equilibrium
conditions which are set by relations among chemical potentials of
the constituents of the neutron star matter. Chemical potentials
depend on the effective baryon masses which have been
substantially modified in the considered nonlinear models. For
comparison the analysis of the influence of the hyperon-hyperon
interaction strength has been included. Dashed lines in presented
figures represent the concentrations
of baryons for the weak $Y-Y$ interaction.\\
 In Fig. \ref{YR0} and \ref{YR008} the analysis of the chemical
composition of the star is depicted. These figures show profiles of
particular baryon and lepton species as a function of the star radius.
One can see that in the case of nonlinear models the hyperon core
spreads through-out almost the whole interior of the star, however,
it includes reduced population of hyperons. It is especially visible
in the case os $\Xi^{-}$ hyperons. For the case $\Lambda_{V}=0.01$,
there is no $\Xi^{-}$ hyperons in the core. This fact ic strictly
connected with the population of leptons, and especially influences
the population of muons. %

\begin{figure}
\begin{centering}
\centering\includegraphics[clip,width=8cm]{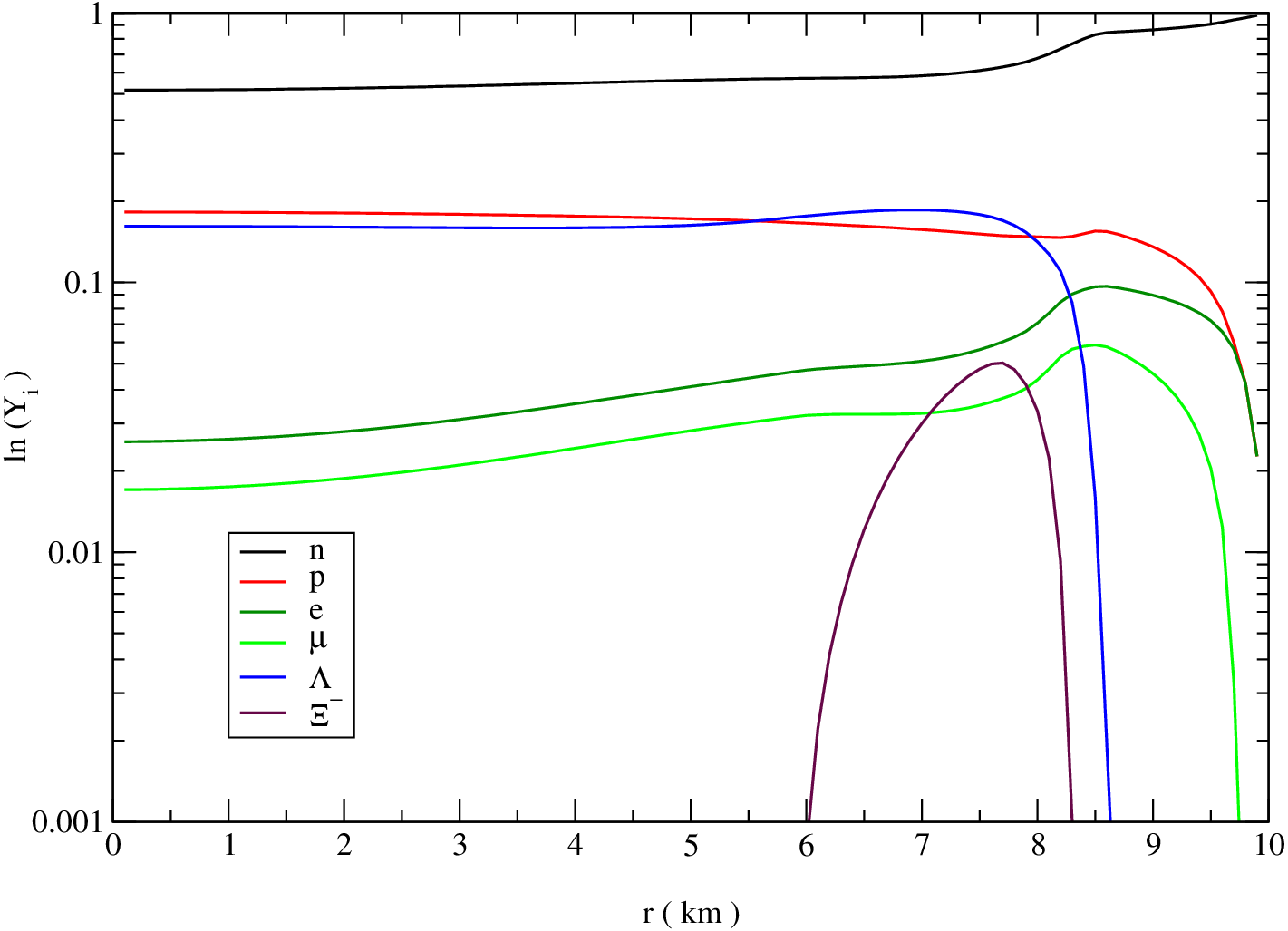}
\par\end{centering}

\caption{Baryon concentrations as a function of stellar radius for $\Lambda_{V}=0.01$}

\label{YR01}
\end{figure}

From the presented figures it is evident that lepton populations are
altered not only by the change of strength of hyperon coupling constants
but also by the isospin dependent nonlinearities which in turn determine
the symmetry energy of the system. Thus a very special aspect of the
existence of hyperons is the intrinsic deleptonization of neutron
star matter. First the appearance of $\Lambda$ hyperons stops the
increase in the lepton population and additionally when negatively
charged $\Xi^{-}$ hyperons emerge further deleptonization occurs.
Thus the charge neutrality can be guaranteed with the reduced lepton
contribution. On the contrary in the hyperon core of of the nonlinear
models when negatively charged $\Xi^{-}$ hyperons disappear lepton
concentrations is substantially enhanced, and large population of
muons establishes. %
\begin{figure}
\begin{centering}
\centering\includegraphics[clip,width=8cm]{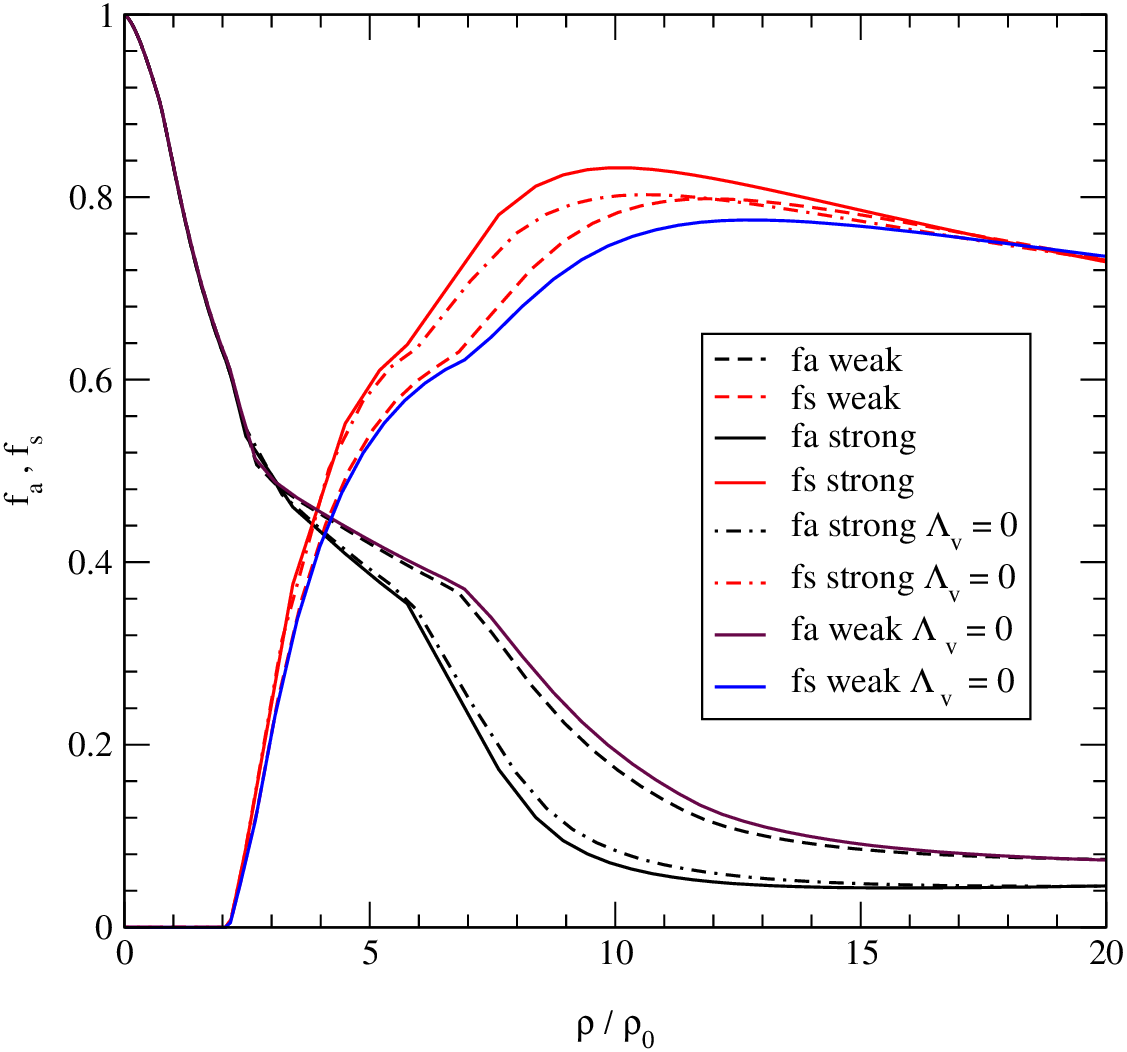}
\par\end{centering}

\caption{The asymmetry parameter $f_a$ and the strangeness
fraction $f_s$ for models with $\Lambda_{V}=0$ as a function of
the baryon number density.}

\label{fafsl}
\end{figure}

\begin{figure}
\begin{centering}
\centering\includegraphics[clip,width=8cm]{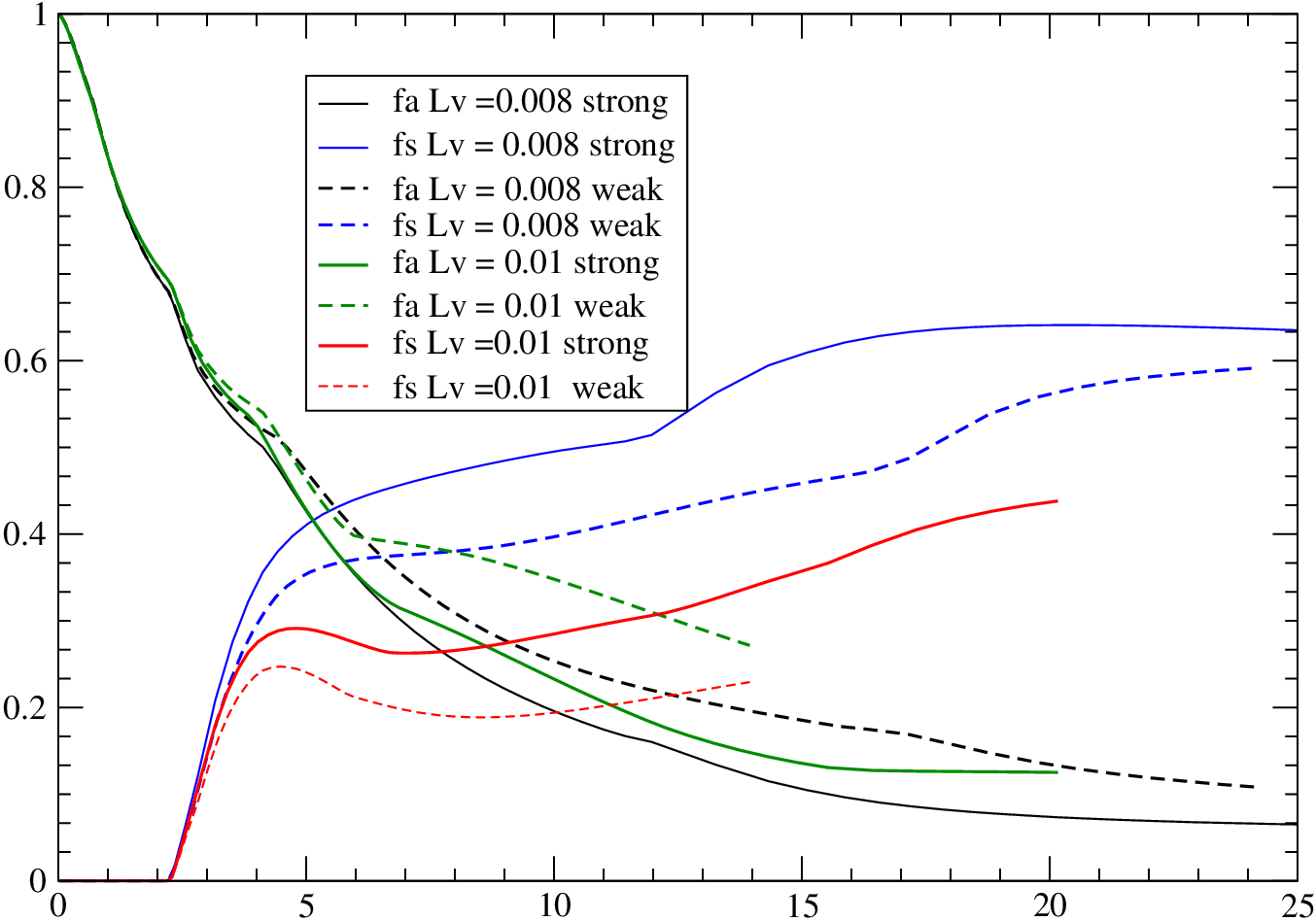}
\par\end{centering}

\caption{The asymmetry parameter $f_a$ and the strangeness
fraction $f_s$ for nonlinear models as a function of baryon number
density.}

\label{fafsnl}
\end{figure}

In Fig \ref{fafsl} and \ref{fafsnl} the density dependence of the
asymmetry parameter $f_{a}=(n_{n}-n_{p})/(n_{b})$ which describes
the relative neutron excess and the strangeness content $f_{s}$,
defined as $(n_{\Lambda}+2n_{\Xi^{-}}+2n_{\Xi^{0}})/n_{B}$ are
presented. Fig. \ref{fafsl} is constructed for the TM 1 parameter
set supplemented by the strange sector and for the case when the
coupling $\Lambda_{V}=0$. The second figure depicts the
differences in the $f_{a}$ and $f_{s}$ behavior for two values of
the $\Lambda_{V}$ coupling. In the case that no nonlinearities are
included the parameter $f_{s}$ takes the maximal value. Next the
nonlinearities have been added with the result that the
strangeness content of the system is reduced. The increase of the
$\Lambda_{V}$ parameter leads to even lower value of the
strangeness fraction. The comparison of these two figure leads to
the conclusion that the nonlinear model include matter more
asymmetric but with lower
value of the strangeness content. %
\begin{figure}
\begin{centering}
\centering\includegraphics[clip,width=8cm]{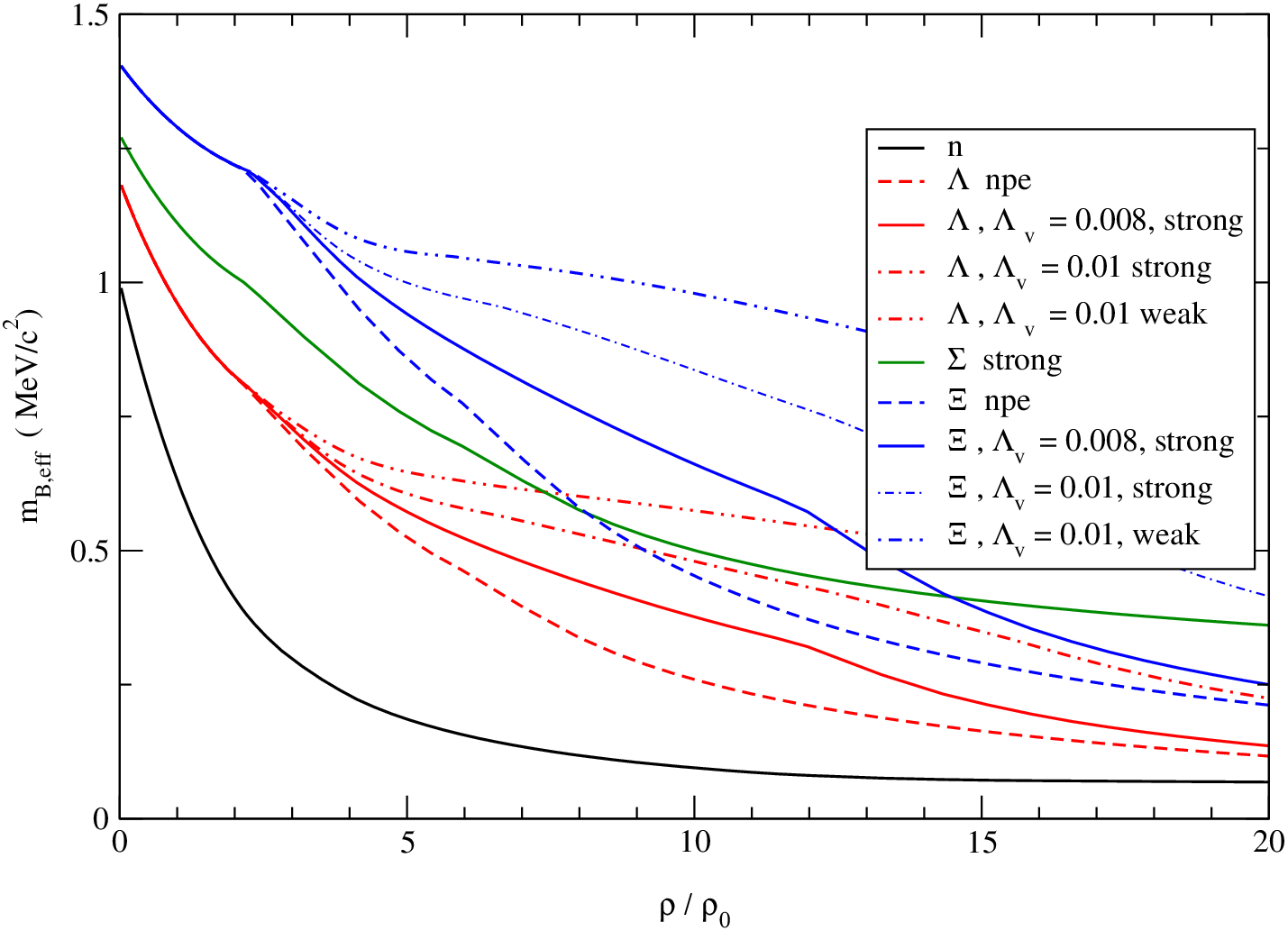}
\par\end{centering}

\caption{The effective baryon masses for different value of the
parameter $\Lambda_{V}$ as a function of baryon number density.}

\label{masy}
\end{figure}

\begin{figure}
\begin{centering}
\centering\includegraphics[clip,width=8cm]{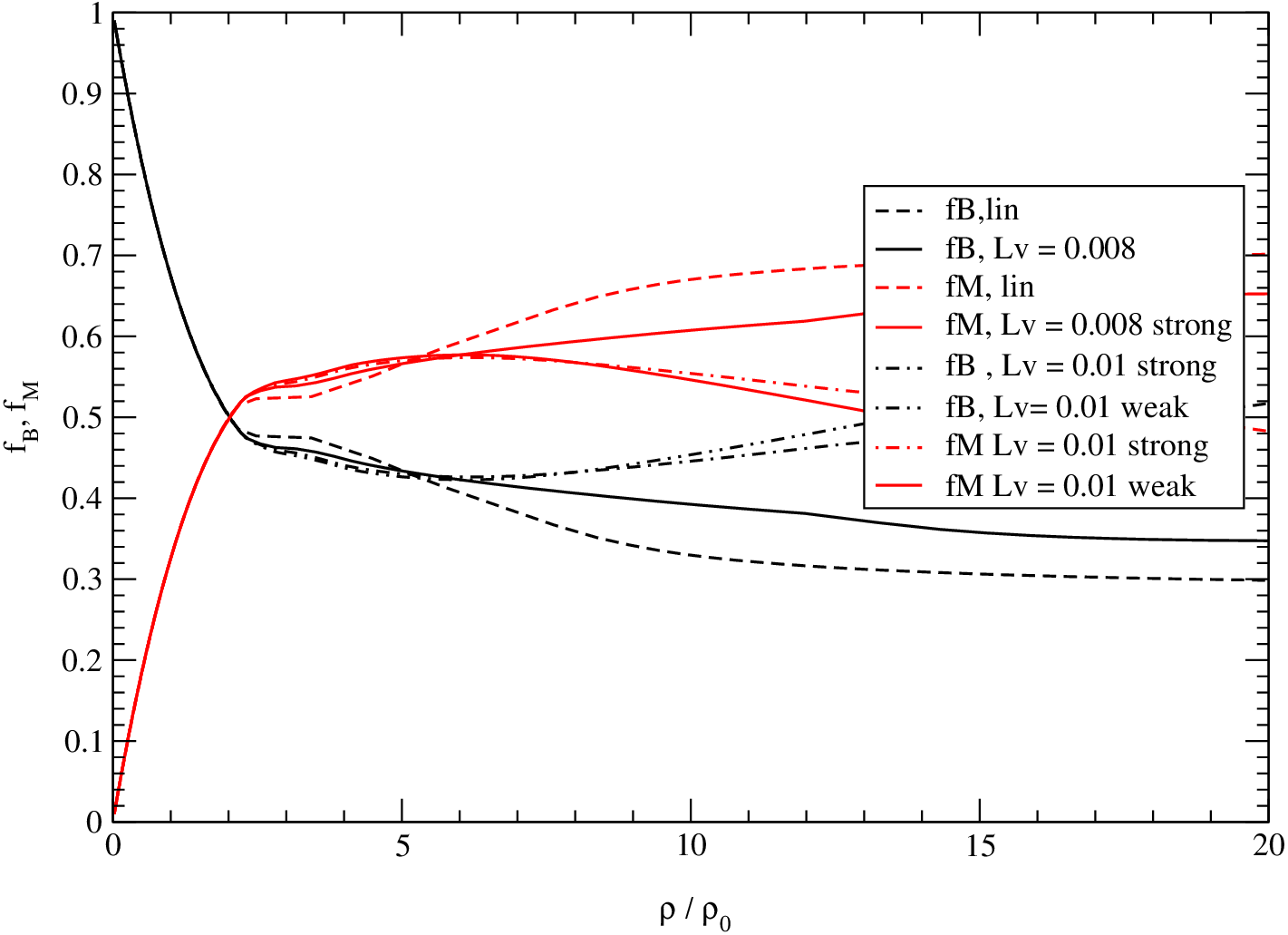}
\par\end{centering}

\caption{The fraction of baryon and meson contribution in the total energy
of the system. Calculations have been done for linear and nonlinear
mdels.}

\label{fBfM}
\end{figure}

Fig. \ref{masy} presents changes of the effective baryon masses in
dependance of the baryon density. It is common feature of the models
that the effective baryon masses decreases with the increasing density.
Very important aspects of the inclusion of nonlinear terms is connected
with change of the effective baryon masses. In general the effective
baryon masses are functions of $s_{0}$ and $s_{0}^{\ast}$. The numerical
calculations show that the inclusion of nonlinear terms leads to the
increase of the effective masses. Analysis has been done in comparison
to the case when no nonlinearites are included. \\
 In Fig. \ref{fBfM} the contributions of mesonic and baryonic
parts to the total energy density is depicted. For the increasing
density the nonlinear models lead to the situation when baryons and
mesons give equal contributions to the energy of the whole system.

\section{Conclusions.}

\begin{figure}
\begin{centering}
\includegraphics[width=9cm]{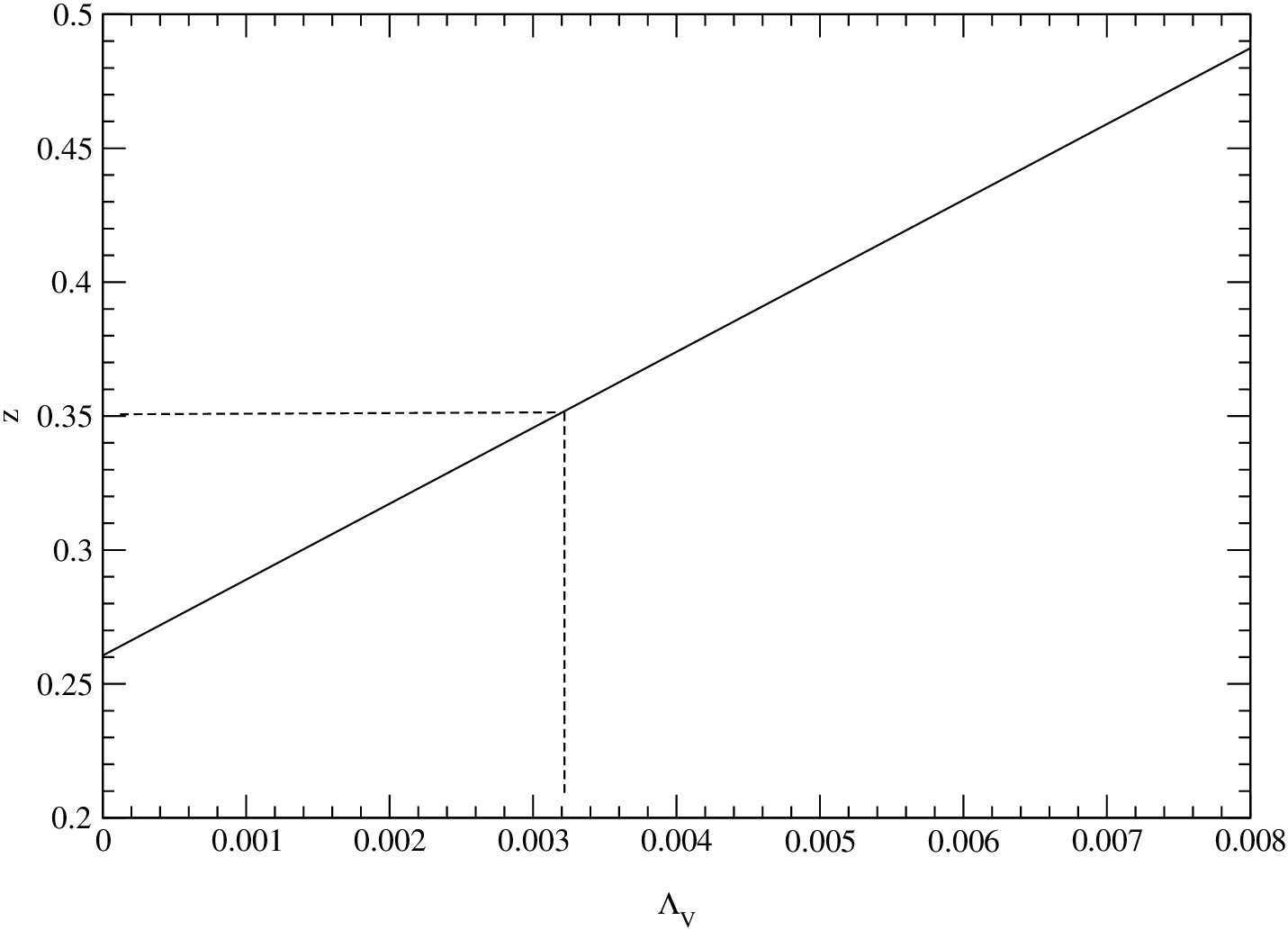} 

\par\end{centering}

\caption{The dependance of the redshift $z$ on the value of the parameter
$\Lambda_{V}$}

\label{rrgLv}
\end{figure}

Detailed knowledge about isospin asymmetric nuclear matter is of
fundamental importance for understanding the structure of a
neutron star, whose formation is preceded by the phenomenon of
supernova explosion. In this paper a special class of the
equations of state of asymmetric nuclear matter with non-zero
strangeness have been analyzed in a systematic approach within the
relativistic mean field model. The basic characteristic of the
considered equations of state is the extended vector meson sector
of the theory. This results in the appearance of various nonlinear
vector meson couplings, among which there are terms which relate
the strange and non-strange mesons. As a consequence strong
connections between the asymmetry and strangeness fraction of the
model have emerged. In order to construct neutron star models for
the obtained equations of state the parameter sets which stem from
the effective field theory and chiral SU(3) theory have been used.
It has been shown that neutron star properties and through the
properties also the neutron star structure are significantly
affected not only by the presence of hyperons but also by the
strength of hyperon-nucleon and hyperon-hyperon interactions. The
results of the analysis performed for the nonlinear models have
been compared with those obtained with the use of the standard TM1
parameter set extended by nonlinear meson interaction terms, which
have been added for detailed investigations of the high density
symmetry energy. It has been shown that in the very nonlinear
models the inclusion of hyperons does not soften the EOS; on the
contrary it leads to its considerable stiffening. The consequences
for neutron star parameters are straightforward and appear as a
considerable growth of neutron star masses. Thus one of the
inevitable conclusions is that in the case of nonlinear models the
inclusion of hyperons does not results in the lowering of a
neutron star mass. This is of special interest when considering
pulsar data reported which indicate large neutron star masses and
radii. This refers to measurements of the neutron star mass in
pulsar-white dwarf system. The analysis performed clearly
indicates that also the structure of a neutron star is changed in
the case of nonlinear models. Stars become more uniform and more
compact. The threshold for the appearance of hyperons is shifted
to the very outer part of the neutron star core, but in general
the hyperon fraction is reduced when compared with the linear
models. The reduction of the hyperon population in neutron star
matter is related to the lepton concentrations. The models show
that for a sufficiently large value of the parameter $\Lambda_{V}$
there are no $\Xi^{-}$ hyperons in the innermost part of the
neutron star. Such particular models of the neutron star core
reveal a hyperon reach layer with $\Lambda$ and $\Xi^{-}$ hyperons
and the central region in which $\Xi^{-}$ hyperons vanish. In the
latter case the populations of leptons and
especially muons get enhanced. \\
 The models analyzed include different types of nonlinear vector
meson couplings. The additional coupling constants, which appear
in these models are specified by the value of the parameter
$\Lambda_{V}$ which is constrained by the requirement of causality
for neutron star matter. The value of $\Lambda_{V}$ has profound
influence on neutron star parameters. This is shown by Fig.
\ref{rrgLv} which depicts the dependance of the redshift $z$ on
the value of the parameter $\Lambda_{V}$. The presented models can
be compare with those which do not include nonlinear vector meson
interactions \cite{04:Hanauske}, these solutions lead to very
particular form of EOS This very particular form of the EOS
generate different solutions of the Oppenheimer-Tolman-Volkoff
equations. In the case of the cold neutron star model, apart from
the ordinary neutron star branch, there exists an additional
stable branch of solutions.

The detection of redshifted O and Fe lines by XMM-Newton from the
surface of the neutron star EXO 0748-676 \cite{2006:Ozel}
indicates for rather stiff EOS.

\end{document}